\documentclass[twocolumn,showpacs,preprintnumbers,amssymb,prd,nofootinbib,aps]
{revtex4}

\usepackage{amsmath}
\usepackage{amsfonts}
\usepackage{amssymb}
\usepackage{amsthm}
\usepackage{graphicx,epsfig,amssymb}
\usepackage{longtable}
\usepackage{bm}

\def\be{\begin{equation}}
\def\ee{\end{equation}}
\def\beq{\begin{eqnarray}}
\def\eeq{\end{eqnarray}}
\def\non{\nonumber}

\def\Lie{\mathcal{L}}
\def\p{\partial}

\newlength{\sizeonefig}
\newlength{\sizetwofig}
\begin{document}

\title{Stability of the puncture method with a generalized 
BSSN formulation}

\author{Helvi Witek$^{1,2}$\footnote{helvi.witek@ist.utl.pt}, 
David Hilditch$^2$\footnote{david.hilditch@uni-jena.de},
Ulrich Sperhake$^{2,3,4,5}$\footnote{sperhake@tapir.caltech.edu}} 
\affiliation{${}^1$ 
  Centro Multidisciplinar de Astrof\'\i sica --- CENTRA, 
  Departamento de F\'\i sica, Instituto Superior T\'ecnico --- IST \\
  Universidade T\'ecnica de Lisboa - UTL,
  Av. Rovisco Pais 1, 1049-001 Lisboa, Portugal}
\affiliation{${}^2$ 
  Theoretisch-Physikalisches Institut,
  Friedrich-Schiller Universit\"at Jena, 
  Max-Wien Platz 1, 07743 Jena, Germany}
\affiliation{${}^3$ 
  Institut de Ci\`encies de l'Espai (CSIC-IEEC), 
  Facultat de Ci\`encies, Campus UAB, E-08193 Bellaterra, Spain}
\affiliation{${}^4$ 
  Theoretical Astrophysics 350-17, 
  California Institute of Technology, Pasadena, CA 91125}
\affiliation{${}^5$ 
  Department of Physics and Astronomy,
  The University of Mississippi, University, MS 38677, USA}

\begin{abstract}
The puncture method for dealing with black holes in the 
numerical simulation of vacuum spacetimes is remarkably 
successful when combined with the BSSN formulation of the 
Einstein equations. We examine a generalized class of
formulations modelled along the lines of the Laguna-Shoemaker
system and including BSSN as a special case.
The formulation is a two parameter generalization of the choice 
of variables used in standard BSSN evolutions. Numerical stability 
of the standard finite difference methods is proven for the 
formulation in the linear regime around flat space, a special 
case of which is the numerical stability of BSSN. Numerical 
evolutions are presented and compared with a standard BSSN 
implementation. 
Surprisingly, a significant portion of the parameter space yields 
(long-term) stable simulations, including the standard BSSN 
formulation as a special case.
Furthermore, non-standard parameter choices typically result
in smoother behaviour of the evolution variables close to the
puncture.
\end{abstract}


\maketitle

\section{Introduction}
Accelerated bodies generate gravitational waves (GWs) in analogy
to the emission of electromagnetic waves by accelerated charges.
The first direct detection of GWs, expected in the course of the
next decade, will not only provide us with the first strong field
tests of Einstein's general relativity but also open up an entire
new window to the universe. The strongest source of GWs are compact
binary systems involving neutron stars and black holes (BHs).
Such compact objects have been known for a long time to represent
the natural end product of stellar evolution. For instance, stellar-mass
BHs are suspected to be the compact members in X-ray binaries
\cite{McClintock:2003gx}. In addition there
is now strong observational evidence for the existence of
supermassive BHs (SMBHs) at the center
of many if not all galaxies \cite{Rees1984, Ferrarese:2004qr}.
Astrophysical observations in recent decades have thus promoted
BHs from the status of a mathematical curiosity to that
of a key player in many astrophysical processes.

While GW emission from compact objects has been theoretically
predicted for quite a while, the waves' weak interaction makes
their direct observation a daunting task, possible only by using modern
high precision technology. In particular, there exists now an
international network of ground-based laser interferometers
(LIGO \cite{Abbott:2005kq, LIGO}, GEO600 \cite{Hewitson:2007zza, GEO},
VIRGO \cite{Acernese:2008zzf} and TAMA \cite{Ando:2001ej}) operating
at or near design sensitivity. A space-borne interferometer
called LISA \cite{Danzmann:2003tv} is scheduled for launch in about
one decade to supplement such observations with exceptional
accuracy in a lower frequency band.
Still, the understanding of the radiated wave patterns
is crucial for the first detection of GWs as well as for the
interpretation of the measured signal.
Eventually, the community will be
able to gain information about characteristic parameters of the BH
system observed via GWs such as the mass ratio and spins.

The modeling of these binary sources of GWs currently employs a
variety of techniques. The
inspiraling phase of a binary black-hole (BBH) prior to
merger as well as the ringdown phase after the merger
can be modeled accurately by the approximate post-Newtonian
\cite{Blanchet2006} and
perturbation methods \cite{Berti:2009kk}, respectively.
In order to simulate the late inspiral and merger
of a BBH, however, numerical methods are required
to solve the fully non-linear Einstein equations.
A numerical treatment requires us to cast the Einstein equations
into the form of a time evolution system. This is most commonly done
by using the canonical Arnowitt-Deser-Misner ``3+1'' decomposition
\cite{Arnowitt:1962hi}
as further developed by York \cite{York1979}; the 4-dimensional
spacetime is decomposed into a family of 3-dimensional hypersurfaces
labeled by a time coordinate. The geometry of spacetime is determined
by the induced 3-metric $\gamma_{ij}$ on the hypersurfaces and their
extrinsic curvature $K_{ij}$, which describes their embedding. The
coordinates are described by the lapse function $\alpha$ and the shift
vector $\beta^i$. These gauge functions represent the coordinate
freedom of general relativity (GR).

For a long time, numerical methods based on this approach
faced a variety of problems including
the specific formulation
of the evolution equations, suitable coordinate choices and
the treatment of singularities inherent in the spacetimes.
The year 2005 brought about
the eventual breakthrough, the first complete simulations
of a BBH coalescence \cite{Pretorius:2005gq, Baker:2005vv,
Campanelli:2005dd}. The ensuing years have produced a wealth
of results on BBH inspiral pertaining to BH recoil, spin
precession and GW data analysis to name but a few
(see \cite{Pretorius:2007nq, Alcubierre:2008, Hannam:2009rd, Hinder:2010vn,
Centrella:2010mx}
for recent reviews).

The current generation of successful numerical codes can be
divided into two categories. The first class uses the so-called
Generalized Harmonic Gauge (GHG) formulation employed in
Pretorius' original breakthrough. The second type of codes
is commonly referred to as {\it Moving Puncture} codes, the method
underlying the simulations of the Goddard and Brownsville
groups. In spite of the remarkable robustness of both methods,
it is fair to say that our understanding of why these techniques
work so well is limited. The Moving Puncture method
in particular has proven robust in even the most demanding
simulations of BBHs involving nearly critical
spins and velocities close to the speed of light
\cite{Dain:2008ck, Sperhake:2008ga, Shibata:2008rq, Sperhake:2009jz}.
Previous investigations of this method have concentrated on the
structure near the singularity and the impact of gauge conditions
\cite{Hannam:2006vv, Brown:2007tb, Brown:2007nt, Hannam:2008sg,
Brugmann:2009gc}.

The purpose of the present work is to shed additional light on which 
ingredients of the Moving Puncture method make this technique so 
successful. The particular focus of our study is on the underlying 
formulation of the Einstein equations, the 
Baumgarte-Shapiro-Shibata-Nakamura (BSSN) formulation 
\cite{Shibata:1995we,Baumgarte:1998te} as well as a modified version 
thereof modeled along the lines of the alternative evolution system
proposed by Laguna and Shoemaker (LaSh) in 2002 \cite{Laguna:2002zc}.
Such a study is beyond purely academic interest. While the currently 
employed techniques appear to work well for 3+1 dimensional simulations 
in the framework of Einstein's general relativity, there is strong 
motivation to push numerical relativity further. A main target of 
gravitational wave observations is the testing of GR versus 
alternative theories of gravity (see \cite{Will:2005va} for an overview,
\cite{Yunes:2009hc}
for solutions of rotating holes in Chern-Simons modified gravity 
and \cite{Salgado:2008xh} for hyperbolicity studies of scalar tensor 
theories of gravity). 
A further application of numerical relativity, in the context of 
high energy physics, as motivated by so-called TeV gravity scenarios 
\cite{Antoniadis:1990ew,ArkaniHamed:1998rs,Antoniadis:1998ig,
Randall:1999ee,Randall:1999vf},
or by the (conjectured) Anti-de Sitter/Conformal Field theory
(AdS/CFT) correspondence 
\cite{Maldacena:1997re, Gubser:1998bc, Witten:1998qj},
will be the simulation of BHs in higher dimensions  
\cite{Yoshino:2009xp,Shibata:2009ad, Shibata:2010wz, 
Zilhao:2010sr,Witek:2010xi,Sorkin:2009bc,Sorkin:2009wh,
Choptuik:2003qd, Lehner:2010pn,Dennison:2010wd}
or non-asymptotically flat spacetimes 
(see, e.g., \cite{Witek:2010qc} for a recent approach).
An improved understanding of the success of the 3+1 GR techniques will 
be crucial in extending numerical relativity successfully along these 
lines of future research. From a more practical point of view, 
alternative schemes might simply be more efficient and result in 
reduced computational requirements. Unfortunately, we will see further 
below that the LaSh system does not result in faster simulations.

This paper is structured as follows. The formulation of the LaSh 
evolution scheme is presented in Sec.~\ref{sec:formulation}. In 
Sec.~\ref{sec:stabana} well-posedness and numerical stability of 
the LaSh system are studied. The LaSh formulation, implemented as an
extension to the \textsc{Lean} code~\cite{Sperhake:2006cy},
is tested numerically with head-on collision and inspiraling BH 
binaries. 
The numerical results are presented in Sec.~\ref{sec:numerics}. 
Finally Sec.~\ref{sec:conclusion} contains our conclusions.

\section{The LaSh Formulation}
\label{sec:formulation}

\subsection{The ADM equations}
\label{subs:Intro}

Both the BSSN and LaSh systems are typically presented as a 
simultaneous conformal decomposition and readjustment of the 
ADM equations~\cite{Baumgarte:1998te,Beyer:2004sv,
Alcubierre:2002kk}. 
For our purposes such a presentation will not suffice. Instead, the 
addition of definition-differential constraints which alters 
the characteristic structure of the system and guarantees 
well-posedness and the conformal decomposition that changes 
to a convenient form of the evolved variables are considered 
separately.

In any case one must first introduce the ADM system, which 
has the evolution part
\begin{eqnarray}
  \p_t\gamma_{ij}&=&-2\alpha K_{ij}+\Lie_\beta\gamma_{ij},\\
  \p_t K_{ij}&=& -D_iD_j\alpha+\alpha[R_{ij}-2K_{ik}K^k_j+K_{ij}K]
  \nonumber\\
  &&+\Lie_\beta K_{ij}.
\end{eqnarray}
and the physical Hamiltonian and momentum constraints
\begin{align}
H&=R+K^2-K_{ij}K^{ij}=0,\\
M_i&=D_j K^j_i -D_i K=0,
\end{align}
where 
\begin{align}
R_{ij}=\Gamma^k_{\textrm{ }ij,k}-\Gamma^k_{\textrm{ }kj,i}+
\Gamma^k_{\textrm{ }kl}\Gamma^l_{\textrm{ }ij}-
\Gamma^k_{\textrm{ }il}\Gamma^l_{\textrm{ }kj}.
\end{align}
When closed with some gauge choice, the ADM system is 
typically only weakly hyperbolic and thus does not admit a 
well-posed Cauchy problem. The BSSN formulation is one of many 
modifications to the ADM system that can yield a strongly 
(or even symmetric) hyperbolic problem when coupled to some 
gauge~\cite{Nagy:2004td}.

\subsection{BSSN Constraint addition}
\label{sub:Cons_addition}

\paragraph*{Definition-differential constraint:}
We define the differential constraint
\begin{align}
  G_i\equiv f_i-\gamma^{jk}\Big(\gamma_{ij,k}-\frac{1}{3}
  \gamma_{jk,i}\Big)=0\,.
\end{align}
Below it will be seen that this choice naturally 
makes $f_i$ coincide with the relevant BSSN variable. 

\paragraph*{Constraint addition:}
The ADM equations are adjusted to 
\begin{align}
\p_t{\gamma}_{ij}&=\textrm{ADM},\label{BSSNUndecom-g}\\
\p_t{K}_{ij}&=\textrm{ADM}+\alpha G_{(i,j)}-\frac{1}{3}
\alpha\gamma_{ij}\left(H+G_{k}^{,k}\right),\\
\p_t{f}_i &= \p_t\bigg(\gamma^{jk}\gamma_{ij,k}-
\frac{1}{3}\gamma^{jk}\gamma_{jk,i}\bigg)_{\textrm{ADM}}
\non\\
&+2\alpha M_i-2\alpha G^jK_{ij}^{\textrm{\small tf}}
+\Lie_\beta G_i\non\\
&+\gamma_{ij}G^k\p_k\beta^j-\frac{2}{3}G_i\p_j\beta^j\,,
\label{BSSNUndecom-Gamma}
\end{align}
where tf denotes the trace-free part.
The principal part, i.e., highest derivatives of variables 
added correspond exactly to those added in the Nagy-Ortiz-Reula (NOR) formulation \cite{Nagy:2004td}
(with $a=b=1$, $c=d=-1/3$ in the notation of~\cite{Gundlach:2006tw}). 
It is for this reason that Gundlach and Mart\'in-Garc\'ia were able 
to identify the two systems when analyzing the principal 
part~\cite{Gundlach:2006tw}).

\subsection{Conformal decomposition and densitization}
\label{sub:conformal}

\paragraph*{Conformal variables and algebraic constraints:}
The LaSh system \cite{Laguna:2002zc} takes as its evolved 
variables
\begin{align}
\tilde{\gamma}_{ij}&= \gamma^{-\frac{1}{3}}\gamma_{ij},
\label{eqn:LaShvars-1}\\
\chi&= \gamma^{-\frac{1}{3}},\\
\tilde{K}&= \chi^{-\frac{3}{2}n_K}K,\\
\tilde{A}^i{}_j&= \chi^{-\frac{3}{2}n_K} ({K^i}_{j}-
\delta^i{}_{j}K/3),\\
\tilde{\Gamma}^i&= \tilde{\gamma}^{jk}\tilde{\Gamma}^i{}_{jk}=-
\p_j\tilde{\gamma}^{ij}.
\label{eqn:LaShvars-4}
\end{align}
The key difference between LaSh and BSSN is that inside LaSh 
the trace and tracefree parts of the extrinsic curvature are 
densitized. Notice, that we recover the standard BSSN equations for
vanishing densitization parameter.
Note that the definition of $G_i$ gives
\begin{equation}
\label{eq:GammaConstraint}
G_i=f_i-\gamma^{jk}\Big(\gamma_{ij,k}-\frac{1}{3}\gamma_{jk,i}
\Big)=\tilde{\gamma}_{ij}\tilde{\Gamma}^j-\tilde{\gamma}^{jk}\tilde{
\gamma}_{ij,k}.
\end{equation}

\paragraph*{Evolution equations and constraints:}
Taking a time derivative of the definitions, substituting the 
evolution equations and rewriting in terms of the evolved 
variables gives the LaSh equations - up to the algebraic constraints 
$D=\ln(\det\tilde{\gamma})=0$, $S=\tilde{\gamma}_{l[i}
\tilde{A}^l{}_{j]}=0$,
$T=\tilde{\gamma}^{ij}\tilde{A}_{ij}=0$, 
which are assumed to be satisfied exactly.
The unknowns evolve according to
\begin{align}
\p_t\tilde{\gamma}_{ij} &= -2\alpha \chi^{3n_K/2}   \tilde{A}_{ij}
        +\beta^k\tilde{\gamma}_{ij,k}+2\tilde{\gamma}_{k(i}\beta^k_{,j)}
        \non\\
        &-\frac{2}{3}\tilde{\gamma}_{ij}\beta^k_{,k}\label{LaSh-g}\\
\p_t\chi &=   \beta^i\chi_{,i} 
            + \frac{2}{3}\chi(\alpha\chi^{\frac{3}{2}n_K}\tilde{K}
            - {\beta^i}_{,i}),\\
\p_t\tilde{A}^i{}_j &= \chi^{-3n_K/2}[-D^iD_j\alpha 
          + \alpha R^i{}_{j}]^{\textrm{\textrm{\small tf}}}\nonumber\\
        & + (1-n_K)\chi^{3n_K/2}\alpha\tilde{K}\tilde{A}^i{}_j
          + \beta^k\tilde{A}^i{}_{j,k}\nonumber\\
        & - \tilde{A}^k{}_j\beta^i_{,k} + \tilde{A}^i{}_k\beta^k_{,j}
          + n_K\tilde{A}^i{}_j\beta^k_{,k}\,,\\
\p_t\tilde{K} &= -\chi^{-3/2n_K} D_iD^i\alpha
          + \beta^k\tilde{K}_{,k} + n_K\tilde{K}\beta^k_{,k} \nonumber\\
        & + \chi^{3n_K/2}\alpha (\tilde{A}^{ij}\tilde{A}_{ij} 
          + (1-3n_K)\tilde{K}^2/3 ),\\
\p_t\tilde{\Gamma}^i &= -2\chi^{3n_K/2} \tilde{A}^{ij}\alpha_{,j}
          + 2\alpha ( \chi^{3n_K/2}\tilde{\Gamma}^i_{jk}\tilde{A}^{jk}
           \nonumber\\
        & -\frac{3}{2}\chi^{3n_K/2} \tilde{A}^{ij}\ln(\chi)_{,j}
          - \frac{2}{3}\tilde{\gamma}^{ij}(\chi^{3n_K/2}\tilde{K})_{,j}) 
             \nonumber\\
        & + \tilde{\gamma}^{jk}\beta^i_{,jk}
          + \frac{1}{3}\tilde{\gamma}^{ij}\beta^k_{,kj}
          + \beta^j\tilde{\Gamma}^i_{,j}
          - \left(\tilde{\Gamma}\right)_{\textrm{d}}^j\beta^i_{,j}\non\\
        & + \frac{2}{3}\left(\tilde{\Gamma}\right)_{\textrm{d}}^i\beta^j_{,j},
\label{LaSh-Gamma}
\end{align}
where $\big($ $\big)_{\textrm{d}}$ denotes the definition of 
those terms rather than the evolved variable and $R_{ij}$ 
is partially rewritten in terms of $\tilde{\Gamma}^i$,
\begin{align}
R_{ij}&= R^\chi_{ij}+\tilde{R}_{ij},\\
R^{\chi}_{ij}& = \frac{1}{2\chi}\tilde{D}_i\tilde{D}_j\chi+\frac{1}{2\chi}
\tilde{\gamma}_{ij}\tilde{D}^l\tilde{D}_l\chi\nonumber\\
&-\frac{1}{4\chi^2}\tilde{D}_i\chi\tilde{D}_j\chi-\frac{3}{4\chi^2}
\tilde{\gamma}_{ij}\tilde{D}^l\chi\tilde{D}_l\chi\\
\tilde{R}_{ij}&= - \frac{1}{2}\tilde{\gamma}^{lm}
\tilde{\gamma}_{ij,lm} +\tilde{\gamma}_{k(i|}\tilde{\Gamma}
^k_{|,j)}+\left(\tilde{\Gamma}\right)_{\textrm{d}}^k
\tilde{\Gamma}_{(ij)k}+\nonumber\\
&\tilde{\gamma}^{lm}\left(2\tilde{\Gamma}^k_
{l(i}\tilde{\Gamma}_{j)km}+\tilde{\Gamma}^k_{im}
\tilde{\Gamma}_{klj}\right).
\end{align}
$\tilde{D}_i$ denotes the covariant derivative compatible with 
the conformal metric. The physical constraints are rewritten
\begin{align}
H&=R-\chi^{3n_K}(\tilde{A}^j{}_k\tilde{A}^k{}_j-
\frac{2}{3}\tilde{K}^2)=0,\label{eqn:LaShHam}\\
M_i&=\tilde{A}^j{}_{i,j}-\frac{2}{3}\tilde{K}_{,i}-\frac{n_K}{\chi}
\tilde{K}\chi_{,i}-\tilde{A}^j{}_m\tilde{\Gamma}^m_{ji}\non\\
&-\frac{3}{2\chi}(1-n_K)\tilde{A}^m{}_i\chi_{,m}=0.\label{eqn:LaShMom}
\end{align}
The differential constraints are given by Eq.~\ref{eq:GammaConstraint}
and algebraic constraints by 
\begin{align}
S&\equiv\tilde{\gamma}_{l[i} \tilde{A}^l{}_{j]}=0\,,\quad &
T\equiv\tilde{\gamma}^{ij}\tilde{A}_{ij}=0\,,\\
D&\equiv\ln(\det\tilde{\gamma})=0.&
\end{align}
Numerical relativity codes use a technique called constraint 
projection to enforce the algebraic constraints. When 
operations are performed which may violate the $D$, $S$ and 
$T$ constraints they are enforced explicitly. It is for this 
reason that we need not worry about the algebraic constraints 
in the construction of (\ref{LaSh-g}-\ref{LaSh-Gamma}); the 
continuum system they represent is identical to that of 
(\ref{BSSNUndecom-g}-\ref{BSSNUndecom-Gamma}). BSSN evolves 
$\tilde{A}_{ij}$, so does not have the symmetry constraint $S$.

\subsection{Gauge conditions}
\label{sub:Gauge}

The successful evolution of binary BH systems has been 
possible with the now standard 1+log variant of the 
Bona-Mass\'o slicing condition, 
\begin{equation}
\p_t{\alpha}=-2\alpha K+\beta^i\partial_i \alpha.\label{eqn:1+log}
\end{equation}
Stationary data for this gauge has been studied in 
\cite{Hannam:2006vv, Brown:2007tb, Brown:2007nt, Hannam:2008sg,
Brugmann:2009gc,Garfinkle:2007yt}. In our numerical evolutions 
the $\Gamma$-driver shift condition  
\begin{align}
  \p_t\beta^i&=\mu_SB^i+\xi_1\beta^j\partial_j \beta^i{},
      \label{eqn:LaShshiftch1}\\
  \p_tB^i&=\p_t\tilde{\Gamma}^i-\xi_2\beta^j\partial_j \tilde{\Gamma}^i{}
      -\eta B^i +\xi_1\beta^j\partial_j B^i,
\label{eqn:LaShshiftch2}
\end{align}
is used with $(\mu_S,\xi_1,\xi_2,\eta)=(1,0,0,1)$ unless otherwise 
stated. We refer to the combination of the 1+log lapse and 
$\Gamma$-driver shift as ``puncture gauge''. Conditions 
in which the lapse and shift are promoted to the status of 
evolved variables are often called {\it live gauge} conditions.
In analytic studies however it is common to consider a fixed, 
densitized lapse 
\begin{equation}
\label{eq:denslapse}
Q=\gamma^{-\frac{n_Q}{2}}\alpha = \chi^{\frac{3n_Q}{2}}\alpha.
\end{equation}
and fixed shift. 
In contrast to BSSN the LaSh system takes the densitized lapse 
$Q$ as a dynamical variable. It is evolved according to the 1+log 
condition (\ref{eqn:1+log}) rewritten in terms of $Q$.
The original LaSh system \cite{Laguna:2002zc} is modified here by 
the consideration of different densitization parameters $n_K$ and $n_Q$ 
for the extrinsic curvature and the lapse. When comparing the 
computational cost of BSSN and LaSh in section~\ref{sec:numerics} 
we additionally evolve LaSh in the downstairs form of the conformal 
extrinsic curvature $\tilde{A}_{ij}$. Whereas in~\cite{Bernuzzi:2009ex} 
the focus was on changing the partial differential equation properties of the formulation and 
holding the variables fixed, here we consider the effect of a change 
of variables alone.

\section{Well-posedness and numerical stability}
\label{sec:stabana}

The well-posedness of the LaSh system with either the puncture 
gauge or a fixed densitized lapse and shift 
(assuming that the $D$,$S$ and $T$ constraints are satisfied) 
was previously studied in~\cite{Gundlach:2006tw,Beyer:2004sv}, so 
verifying these properties for the system linearized around flat 
space is in its own right uninteresting. However we wish to 
demonstrate the numerical stability of the LaSh system around 
flat-space. The approach for the semi-discrete scheme is analogous 
to that for the continuum system, so we first tackle that problem.
In Sec.~\ref{sect:Theoretical} we briefly recap the 
theoretical background. Next, in Sec.~\ref{sect:Cont_Puncture} 
we use characteristic variables to demonstrate well-posedness 
for the continuum system. The analysis is then extended to the 
semi-discrete case, and follows closely the method of 
\cite{Calabrese:2005ft,Chirvasa:2008xx}. Finally, we deal with 
the algebraic constraints in the fully-discrete system by 
demonstrating that the natural semi-discrete limit of the 
standard implementation (with constraint projection) is given 
by the systems considered in Sec~\ref{sect:Cont_Puncture}.

\subsection{Theoretical background }
\label{sect:Theoretical}

\paragraph*{Continuum system:} The linear, constant coefficient,
first order in time and second order in space time evolution 
problem 
\begin{align}
\p_tu=P[\p_x]u,\quad u(t=0,x)=f(x)
\end{align}
is called well-posed with respect to a norm $||\cdot||$ if for 
every smooth, periodic $f(x)$ there exists a unique smooth 
spatially periodic solution and there are constants $C, K$ such 
that for $t\geq 0$ 
\begin{align}
||u(t,\cdot)|| \le Ke^{C t}||u(0,\cdot)||.
\end{align}
A hermitian matrix $\hat{H}(\omega)$ is called a symmetrizer of the 
system if the energy $\hat{u}^*\hat{H}\hat{u}$ is conserved by 
the principal part of the Fourier transformed system, with 
\begin{align}
K^{-1}I_\omega\leq \hat{H}\leq K I_\omega,\quad 
I_\omega\equiv\left(\begin{array}{cc}
\omega^2 & 0 \\
0 & I \end{array}\right),
\end{align}
for some $K>0$ constant, for every frequency $\omega$ in Fourier 
space ($\hat{u}$ denotes the Fourier transformed function.)
We say that the Hermitian matrices $A,B$ satisfy the 
inequality $A\le B$ if $y^\dagger A y \le y^\dagger B y$ for every
$y$. Well-posedness is equivalent to the existence of a 
symmetrizer~\cite{Gustafsson1995}, which is in turn equivalent
to the existence of a complete set of characteristic variables.

\paragraph*{Discrete system:}
We introduce a grid 
\begin{align}
\mathbf{x}_j=(x_{j_1},y_{j_2},z_{j_3})=(j_1h,j_2h,j_3h), 
\end{align}
with
$j_i=0,\dots N_r-1$ and $h=\frac{2\pi}{N}$ is the spatial resolution. 
We denote 
\begin{align}
D_+v_i&=\frac{1}{h}(v_{i+1}-v_i),\quad &D_-v_i= \frac{1}{h}
(v_{i}-v_{i-1}),\\
D_{0}v_i&=\frac{1}{2h}(v_{j+1}-v_{j-1}).&
\end{align}
The standard second order accurate discretization is written
\begin{align}
\p_i &\to D_{0i},\\
\quad\p_i\p_j&\to D^{(2)}_{ij}=\left\{\begin{array}{ll}
D_{0i}D_{0j} & i\neq j\\
D_{+i}D_{-i} & i=j\end{array}\right.
\end{align}
For brevity, we do not consider higher order discretization. Fourier 
transforming reveals
\begin{align}
\hat{D}_i^{(1)}&=\frac{i}{h}s_i,\\
\hat{D}_{ij}^{(2)}&=
\left\{\begin{array}{ll}
-\frac{1}{h^2}s_is_j & i\neq j\\
-\frac{4}{h^2}t_i^2t_j^2 & i=j
\end{array}\right.
\end{align}
We abbreviate $s_i = \sin\xi_i$ and $t_i = \sin\xi_i/2$, and write
\begin{align}
\xi_r & = \omega_r h = -\pi + \frac{2\pi}{N}, -\pi + \frac{4\pi}{N}, 
... ,\pi.
\end{align}
where $\omega_r=-\frac{N}{2}+1,...,\frac{N}{2}$ and $r=1,2,3$.
The time step $k$ is related to the spatial resolution $h$  through 
the Courant factor $k=\lambda_ch$. Finally, we use the notation
\begin{align}
\omega^2= -\eta^{ij}\hat{D}_{0i}\hat{D}_{0j},&\qquad
\Omega^2= -\eta^{ij}\hat{D}^{(2)}_{ij},
\end{align}
where $\eta^{ij}$ is just the identity matrix, so that 
$\Omega^2 = \sum_{i=1}^3 |\hat{D}_{+i}|^2$. The results for numerical 
stability with a polynomial method of lines time-integrator 
are analogous to the result at the continuum: if there exists a 
hermitian $\hat{H}(\xi)$ for every grid frequency $\xi$ such that the 
energy $\hat{u}^*\hat{H}\hat{u}$ is conserved by the Fourier-transformed 
semi-discrete principal system and satisfies
\begin{align}
K^{-1}I_{\Omega}\le\hat{H}\le K I_{\Omega},\quad 
I_\Omega\equiv\left(\begin{array}{cc}
\Omega^2 & 0 \\
0 & I \end{array}\right),
\end{align}
with $K$ as above, then it is possible to construct a discrete 
symmetrizer for the semi-discrete problem without lower 
order terms. If the spectral radius of the product of the time-step 
and the semi-discrete symbol is bounded by a value that depends on the 
time-integrator, then the system is stable with respect to the norm
\begin{align}
||u||^2_{h,D_+}\equiv ||u_{s}||^2_{h}+
||u_{f}||^2_{h}+\sum_{i=1}^3||D_{+i}u_{s}||^2_{h},
\end{align}
where the subscript distinguishes between variables that appear as 
second derivatives in the continuum system. The estimate
\begin{align} 
||u^{n\Delta t}||_{h,D_+}\le Ke^{C n\Delta t}||u^0||_{h,D_+},
\end{align}
then holds. For details we refer the reader to
\cite{Calabrese:2005ft}.

\paragraph*{Discussion:} A straightforward way to construct 
characteristic variables for the continuum system is to perform a 
$2+1$ decomposition. One then ends up with decoupled 
scalar, vector and tensor sectors which are hopefully straightforward 
to diagonalize. Diagonalisability of a system guarantees the 
existence of a complete set of eigenvectors, which in turn guarantees 
well-posedness in some norm. For the discrete system, a similar 
approach is not possible with the standard discretization because 
the various blocks of the system remain coupled. This complication
is caused by the fact that under the standard discretization the 
second derivative is not equivalent to a repeated application of 
the first derivative. We will see in the following sections that 
this forces us to consider significantly larger matrices, 
and that for the main case of interest, the stability of the LaSh 
formulation with puncture gauge, the calculation is impractical.
One may also consider the numerical stability of systems with 
the $D_0^2$ discretization, in which second derivatives are 
approximated by repeated application of the centered difference 
operator $D_0$. In this case it is possible to make a $2+1$ 
decomposition of the semi-discrete system. Unfortunately, the 
discretization suffers from the problem that the highest 
frequency mode on the grid is not captured by the scheme. For
the Fourier transformed system this property implies that the 
transformed spatial derivatives vanish, which typically prevents 
one from building an estimate on the highest frequency mode. 
Although artificial dissipation may restore stability, we do not 
consider the $D_0^2$ discretization further.

\subsection{Continuum system}
\label{sect:Cont_Puncture}

\paragraph*{Fixed densitized lapse and shift:} We begin by 
linearizing the LaSh system around flat-space. Following 
\cite{Gustafsson1995,Calabrese:2005ft} we Fourier transform in 
space, and make a pseudo-differential reduction to first order. 
Spatial derivatives transform according to $\p_i \rightarrow 
\imath w_i$. The system has a complete set of characteristic 
variables with characteristic speeds 
$(0,\pm\omega,\pm\sqrt{n_Q}\omega)$. A conserved quantity for 
the system may be trivially constructed from the characteristic 
variables. It is straightforward but tedious to demonstrate that 
the conserved quantity is equivalent to the norm
\begin{equation}
||u||_{\textrm{\small{fd}}}^2=||\gamma_{ij}||^2+||K_{ij}||^2+||f_i||^2
+\sum_{k=1}^{3}||\gamma_{ij,k}||^2
\end{equation}

\paragraph*{Puncture gauge:}
For simplicity we consider the time-integrated $\Gamma$-driver 
shift condition
\begin{align}
\p_t\beta_i=f_i.
\end{align}
The transformed vector of evolved variables is 
$\hat{u}=(\hat{\gamma}_{ij},
\hat{\alpha},\hat{f}_k,\hat{K}_{lm},\hat{\beta}_n)$. 
The principal symbol is
\begin{widetext}
\begin{align}
{\hat{P}^{\mu}}_\nu=\left(\begin{array}{ccccc}
0 & 0 & 0 & -2\delta^l_i\delta^m_j & 2\imath w\hat{\omega}_{(i}\delta^n_{j)}\\
0 & 0 & 0 & -2\eta^{lm} & 0\\
0 & 0 & 0 & -\frac{4}{3}\imath \eta^{lm} w\hat{\omega}_{k} & 
- w^2 \left(\delta^n_k+\frac{1}{3}\hat{\omega}^n\hat{\omega}_k \right)\\
\frac{1}{2}w^2[\delta^i_l\delta^j_m+\frac{1}{3}\eta^{ij}\hat{\omega}_{l}
\hat{\omega}_m]^{\textrm{{\small tf}}}
& w^2 \hat{\omega}_{l}\hat{\omega}_m 
& \imath w [\hat{\omega}_{(l}\delta^k_{m)}]^{\textrm{{\small tf}}}
& 0 & 0\\
0 & 0 & \delta^k_n & 0 & 0
\end{array}\right).
\end{align}
\end{widetext}
We denote $w_i=w\hat{w}_i$, with $w=|w|$. Here ``trace-free'' denotes 
that the trace is removed in downstairs indices. The characteristic 
variables can be constructed from the matrix
\begin{widetext}
\begin{align}
{\small{
T^{-1}=\left(\begin{array}{ccccc}
-\frac{1}{3}\imath w\eta^{ij} & -\frac{1}{3}\imath w & \hat{\omega}^k & 0 & 0\\
0 & \pm \frac{1}{\sqrt{2}}\imath w & 0 & \eta^{lm} & 0\\
0 & -\imath w &  \hat{\omega}^k & \pm\sqrt{\frac{2}{3}}\eta^{lm} & 
\pm\sqrt{\frac{2}{3}}\imath w\hat{\omega}^n\\
0 & 0 & -\hat{\omega}_{(i}\delta^k_{j)}+\hat{\omega}_{i}\hat{\omega}_{j}
\hat{\omega}^{k} 
& 0 & \pm\imath w[\hat{\omega}_{(i}\delta^n_{j)}-\hat{\omega}_{i}
\hat{\omega}_{j}
\hat{\omega}^{n}]  \\
\frac{1}{2}\imath w[\delta^i_p\delta^j_q+\frac{1}{3}
\hat{\omega}_p\hat{\omega}_q\eta^{ij}]^{\textrm{{\small tf}}}
&
-\frac{8}{3}\imath w[\hat{\omega}_{p}\hat{\omega}_{q}]^
{\textrm{{\small tf}}} & 
-[\delta^k_{(p}\hat{\omega}_{q)}-2\hat{\omega}_p\hat{\omega}_q
\hat{\omega}^k]^{\textrm{{\small tf}}} 
& 
\pm[\delta^l_p\delta^m_q-\frac{11}{3}\eta^{lm}\hat{\omega}_{p}
\hat{\omega}_{q}]^{\textrm{{\small tf}}} 
& 
\pm 2\imath w[\hat{\omega}_{p}\hat{\omega}_{q}]^{\textrm{{\small tf}}}
\hat{\omega}^{n}
\end{array}\right)
}}
\end{align}
\end{widetext}
through $U_c=T^{-1}u$.
The characteristic speeds corresponding to each row are 
$(0,\pm \sqrt{2}w,\pm\sqrt{2/3}w,\pm w,\pm w )$. The conserved 
quantity is given by
\begin{center}
\begin{align}
&   \left(-\frac{4a}{\epsilon_1}
        + f\left( 1-\frac{1}{\epsilon_3} - \frac{4}{\epsilon_5} 
             - \frac{8}{3\epsilon_7} \right)\right)
    \frac{w^2}{4}|\hat{\gamma}_{ij}|^2  \nonumber\\
& + \left(\frac{a}{9} + \frac{c}{2} +d \right) w^2|\hat{\alpha}|^2 
  + e w^2|\hat{\beta}_i|^2 
  + f |\hat{K}_{ij}|^2 \nonumber\\
& + \left(e - a\epsilon_1 
    - f\left(\epsilon_3+\epsilon_5+\frac{2\epsilon_7}{3}\right)\right) 
        |\hat{f}_i|^2 \nonumber\\
& \leq E_C \leq  \\
&   \left(4a\left(\frac{1}{3}+\frac{1}{\epsilon_2} \right) 
        +f\left(\frac{28}{9}+\frac{1}{\epsilon_4}+\frac{4}{\epsilon_6}
           +\frac{8}{3\epsilon_8} \right) \right)
        \frac{w^2}{4}|\hat{\gamma}_{ij}|^2 \nonumber\\
& + \left(\frac{a}{9} + \frac{c}{2} +d + \frac{256}{27}f\right) 
    w^2|\hat{\alpha}|^2 \nonumber\\
& + \left( 2d + 4e + 16f \right) w^2|\hat{\beta}_i|^2 
  + \left( 3c + 2d + \frac{436}{9}f \right) |\hat{K}_{ij}|^2  \nonumber\\
& + \left( a(3+\epsilon_2) + 3d + 4e 
        + f\left(14+\epsilon_4+\epsilon_6+\frac{2\epsilon_8}{3} \right)\right) 
        |\hat{f}_i|^2 \nonumber.
\end{align}
\end{center}
By choosing $a=\frac{1}{24}$, $c=d=f=1$, $e=26$ and 
$\epsilon_1 = 2$, $\epsilon_2 = \epsilon_4 = \epsilon_6 = \epsilon_8 = 1$,
$\epsilon_3 = 4$, $\epsilon_5 = 16$, $\epsilon_7 = 8$
we obtain
\be
K^{-1} ||\hat{u}||_{\textrm{\small{pg}}}^2 \leq E_C \leq 
K ||\hat{u}||_{\textrm{\small{pg}}}^2
\ee
where $K=125$, and have demonstrated that the conserved quantity is 
equivalent to the norm
\be
||\hat{u}||_{\textrm{\small{pg}}}^2 = w^2||\hat{\gamma}_{ij}||^2
+w^2||\hat{\alpha}||^2+w^2||\hat{\beta}_i||^2+||\hat{K}_{ij}||^2+
||\hat{f}_i||^2\,.
\ee
Parseval's relation implies equivalence with 
\begin{align}
||u||_{\textrm{\small{pg}}}^2&=||\gamma_{ij}||^2+||\alpha||^2+||\beta_i||^2
+||K_{ij}||^2+||f_i||^2\nonumber\\
 & +\sum_{k=1}^{3}(||\gamma_{ij,k}||^2+||\alpha_{,k}||^2
+||\beta_{i,k}||^2)
\end{align}
in physical space.

\subsection{Discrete system}\label{sec:StabilityAnaDiscrete}

\paragraph*{Fixed densitized lapse and shift:} We now consider the 
semi-discrete system with fixed densitized lapse and shift. As 
in the continuum case we linearize around flat-space. The 
difference operators transform as described in 
Sec.~\ref{sect:Theoretical}. We consider only the case $n_Q=1$.
We define
\begin{align}
\hat{\tau}=\eta^{ij}\hat{\gamma}_{ij},\quad
\hat{\Gamma}_i= \hat{f}_i-\frac{2}{3}D_{0i}\hat{\tau}.
\end{align}
Decomposing the system into trace, off-diagonal, and diagonal 
terms adjusted by the weighting
$t_i^4\hat{\gamma}_{ii}=\tilde{\gamma}_{ii}$ and
$t_i^4\hat{\Gamma}_{i}=\tilde{\Gamma}_{i}$ various sectors 
of the system decouple. In the following $\gamma_{ij}$ explicitly 
means $i\ne j$. The principal symbol is
\begin{align}
(0)&\quad \hat{\Gamma}_{i},\\
\left(\begin{array}{cc}0 & -2 \\\frac{1}{2}\Omega^2 & 0
\end{array}\right) &\quad \left(\begin{array}{c}
\hat{\tau} \\ \hat{K} \end{array}\right),\\
\left(\begin{array}{ccc}
0 & 0 & -2 \\
0 & 0 & 0 \\ 
\frac{1}{2}\Omega^2 & 1 & 0
\end{array}\right)
&\quad
\left(\begin{array}{c}
\hat{\gamma}_{ij} \\
\hat{D}_{0(i}\hat{\Gamma}_{j)}\\
\hat{K}_{ij}
\end{array}\right),\\
\left(\begin{array}{ccc}
0 & 0 & -2 \\
0 & 0 & 0 \\
\frac{1}{2}\Omega^2 & 1 & 0
\end{array}\right) & \quad
\left(\begin{array}{c}
\tilde{\gamma}_{ii}^{\textrm{{\small tf}}} \\
(\hat{D}_{0i}\tilde{\Gamma}_{i})^{\textrm{{\small tf}}}\\
\tilde{K}_{ii}^{\textrm{{\small tf}}}
\end{array}\right).
\end{align}
The characteristic variables are
\begin{align}
&\hat{\Gamma}_i,\\
&K\pm\frac{\imath\Omega}{2}\tau,\\
&K_{ij}\pm\frac{\imath\Omega}{2}\gamma_{ij}\pm\frac{\imath}{\Omega}
D_{0(i}\Gamma_{j)},\\
&\left(\tilde{K}_{ii}\pm\frac{\imath\Omega}{2}
\tilde{\gamma}_{ii}\pm\frac{\imath}{\Omega}
D_{0i}\tilde{\Gamma}_{i}\right)^{\textrm{{\small tf}}},
\end{align}
and have speeds $(0,\pm\imath\Omega,\pm\imath\Omega,\pm\imath\Omega)$.
The system has a pseudo-discrete reduction to first order that 
admits a symmetrizer for every grid-frequency. One must treat the
lowest frequency separately, but in that case the principal 
symbol vanishes and so admits the identity as a symmetrizer.
By the equivalence of norms in finite dimensional vector spaces 
we then have numerical stability in the pseudo-discrete norm
\begin{align}
||\hat{u}||^2_{h,fd}&=\Omega^2 ||\hat{\gamma}_{ij}||^2_h 
+||\hat{K}_{ij}||^2_h+||\hat{f}_i||^2_h+||\hat{\gamma}_{ij}||^2_h,
\end{align}
provided that the von-Neumann condition given by
\begin{equation}
\lambda_C \leq \frac{C_0}{2\chi_2}\,,
\end{equation}
where $C_0 = 2$ and $C_0 = \sqrt{8}$ for iterated Crank-Nicholson 
or fourth-order Runge-Kutta and $2\chi_2 = \Omega h$,
is satisfied. Parseval's relation 
guarantees equivalence with the discrete norm 
\begin{align}
||u||^2_{h,D_+,fd} &= ||\gamma_{ij}||^2_h +||K_{ij}||^2_h 
+ ||f_i||^2_h + \nonumber\\
&\sum_{k=1}^3||D_{+k}\gamma_{ij}||^2_h
\end{align}
in physical space.

\paragraph*{Puncture gauge:} The principal symbol is a 
$19$x$19$ matrix which contains several parameters. We are 
able to compute characteristic speeds for the system, but they 
are complicated, so we do not display them here. The lowest 
freqency mode is again trivial to analyze. In that case the 
pseudo-discrete reduction to first order has a vanishing
principal symbol, and thus admits the identity as a symmetrizer. For the 
non-maximal modes the principal symbol is complicated. As 
previously stated, performing a $2+1$ decomposition on the 
semi-discrete symbol is not helpful, since the various sectors 
of the system remain coupled. We were therefore unable to find 
the eigenvectors of the matrix.
We considered various subsectors of the full system. Since the ($\alpha$,$K$) 
subsector is exactly the second order in space wave equation, 
it is trivial to demonstrate numerical stability. We considered 
also the subsector ($f_i$,$\beta_j$) with the other variables 
frozen, and find characteristic speeds and variables. Once the 
two blocks are coupled to give the ($\alpha$,$f_i$,$K$,$\beta_j$) 
subsector, we did not manage to compute eigenvectors 
in finite time. It is possible to find a complete set of 
characteristic variables for the highest frequency grid-mode,
since the principal symbol in that case takes a simpler form. In 
Sec.~\ref{sec:numerics} we present robust stability tests of 
the numerical implementation, which provide evidence that the 
system is formally numerically stable.

\paragraph*{Puncture gauge with non-standard spatial 
discretization:} We are able to find a complete set of 
characteristic variables for a slightly altered spatial 
discretization. If one insists on using the $D_0^2$ operator 
for the divergence terms $\p_i\p_j\beta^j$ in the evolution of 
$f_i$ the principal symbol becomes, with 
$\hat{u}=(\hat{\gamma}_{ij},\hat{\alpha},\hat{f}_k,\hat{K}_{lm},
\hat{\beta}_n)$,
\begin{widetext}
\begin{align}
{\hat{P}^{\mu}}_\nu=\left(\begin{array}{ccccc}
0 & 0 & 0 & -2\delta^l_i\delta^m_j & 2\hat{D}_{(0i}\delta^n_{j)}\\
0 & 0 & 0 & -2\eta^{lm} & 0\\
0 & 0 & 0 & -\frac{4}{3} \eta^{lm}\hat{D}_{0k} & 
- \Omega^2\delta^n_k+\frac{1}{3}{\hat{D}_0}^n\hat{D}_{0k}\\
\frac{1}{2}[\Omega^2\delta^i_l\delta^j_m-\frac{1}{3}\eta^{ij}
\hat{D}^{(2)}_{lm}]^{\textrm{{\small tf}}}
& -\hat{D}^{(2)}_{lm}
& [\hat{D}_{0(l}\delta^k_{m)}]^{\textrm{{\small tf}}}
& 0 & 0\\
0 & 0 & \delta^k_n & 0 & 0
\end{array}\right).
\end{align}
\end{widetext}
As before, one has to consider the lowest and highest frequency grid modes 
seperately, because in those cases the principal symbol of the system 
takes a different form. For the lowest frequency mode the principal 
symbol of the pseudo-discrete reduction to first order again vanishes, 
and can be dealt with as before. For the sub-maximal frequencies 
the characteristic variables can be constructed and have characteristic 
speeds $(0,\pm\imath\sqrt{2}\Omega,\pm\imath\sqrt{(3\Omega^2+\omega^2)/3},
\pm \imath\Omega,\pm\imath\Omega)$. The $0$-speed characteristic variable 
is
\begin{align}
U_0&=\hat{D}_{0i}\hat{f}^i+\frac{1}{2}(\omega^2-\Omega^2)\hat{\alpha}
+\frac{1}{6}(3\Omega^2+\omega^2)\tau.
\end{align}
The lapse characteristic variable is
\begin{align}
U_{\pm\sqrt{2}}&=\sqrt{2}\hat{K}\pm\imath\Omega\hat{\alpha},
\end{align}
The longitudinal shift characteristic variable is 
\begin{align}
U_{\pm2/\sqrt{3}}&= \imath\hat{D}_{0i}\beta^i
-\frac{4\imath}{\omega^2-\Omega^2}\hat{K}
\pm\frac{\sqrt{3}}{3\Omega^2+\omega^2}\hat{D}_{0i}f^i\nonumber\\
&\mp\frac{4\sqrt{3}}{\sqrt{3\Omega^2+\omega^2}
(\omega^2-3\Omega^2)}\hat{\alpha},
\end{align}
and the transverse shift modes are 
\begin{align}
U_{i\pm 1}&= \Omega(\hat{D}_{0i}\hat{D}_{0}^k
+\omega^2\delta^k{}_i)\beta_k
\nonumber\\
&\pm(\hat{D}_{0i}\hat{D}_{0}^k+\omega^2\delta^k{}_i)f_k.
\end{align}
To see that there are only two characteristic variables here one must 
contract with the vector $s_i$. Finally the remaining characteristic variables 
are
\begin{align}
U_{ ij \pm 1}& = [\hat{D}^{(2)}_{ij}+
\omega^2\hat{D}_{0(i}\delta^{k}_{j)}]^{{\textrm {\small tf}}}\hat{\beta}_k
\mp 
\frac{\imath}{\Omega}[\hat{D}^{(2)}_{ij}
\hat{D}_{0}^k]^{{\textrm {\small tf}}}\hat{f}_k\nonumber\\  
&+\omega^2[\delta^{k}{}_{(i}\delta^l{}_{j)}
+\eta^{kl}\hat{D}^{(2)}_{ij}]\hat{K}_{kl}\nonumber\\
&\pm\frac{1}{2}\imath\omega^2\Omega[\delta^l{}_{(i}\delta^m{}_{j)}
-\frac{1}{3\Omega^2}\eta^{lm}\hat{D}^{(2)}_{ij}
]^{{\textrm {\small tf}}}\hat{\gamma}_{lm}
\end{align}
where $[$~$]^{{\textrm {\small tf}}}$ denotes that the object is trace-free in 
downstairs indices. For the highest frequency mode the $D_0$ operator in the 
principal symbol vanishes. However, the symbol still has a complete set of 
characteristic variables. The conserved quantity may then be constructed as 
before with a sum over the grid-modes. The conserved quantity is obviously 
a norm since it contains every grid-mode, and is therefore equivalent to 
the standard
\begin{align}
||u||_{h,D_+,\textrm{\small{pg}}}^2&=||\gamma_{ij}||^2_h
+||\alpha||^2_h+||\beta_i||_h^2
+||K_{ij}||^2_h\nonumber\\
&+||f_i||^2_h+\sum_{k=1}^{3}(||D_{+k}\gamma_{ij}||_h^2+
||D_{+k}\alpha||_h^2\nonumber\\
&+||D_{+k}\beta_{i}||_h^2)
\end{align}
in physical space.

As the calculation does not rely in any significant way on the flat
background, it should be simple to extend to the case in which one 
linearizes around an arbitrary, constant in space background.
It may then be possible to extend to the case with variable 
coefficients in space following \cite{Gustafsson1995}. We also 
anticipate no problems in extending the calculation to higher 
order finite difference (FD) approximations.
In our numerical tests in Sec.~\ref{sec:numerics} we do 
not perform evolutions with this discretization. 

\subsection{The algebraic constraints}

In this section we demonstrate that the numerical stability 
of standard numerical implementation of the linearized LaSh 
(and BSSN) systems, which includes the conformal decomposition 
of the evolved variables, depends only upon the analysis of the 
previous section. In order to do so, we show that there is a 
one-one correspondence between solutions of the original and 
decomposed systems.

In the linear regime the conformal decomposition is simply a
linear combination of the undecomposed variables subject to 
linear constraints. Consider the semi-discrete system under 
such a decomposition. Start by defining the decomposed state 
vector on a time slice by $v=Tu$. Here and in what follows we 
suppress spatial indices. Assume that~$u$ has~$m$ elements. 
Then $T$ is an $l\times m$ matrix, $v$ has $l$ elements. We 
denote the pseudo-inverse of $T$ by $S$, a matrix which maps 
from the image of $T$ in ${\mathbf R}^l$ back to 
${\mathbf R}^m$ such that 
\begin{align}
ST=I_{m}.
\end{align}
If the evolution equations for $u$ are given by $Pu$ and those 
for the decomposed variables are $\bar{P}v$ then the two are 
related as $\bar{P}=TPS$. Denote by $\perp$ the projection 
operator which maps to the $m$ dimensional hypersurface in 
${\mathbf R}^l$ on which the algebraic constraints are satisfied.
The algebraic constraints are 
\begin{align}
C&=v-\perp v.
\end{align}

Consider first the semi-discrete system. Suppose that at a given 
time the constraints are satisfied. Then we find
\begin{align}
\p_tC&=\bar{P}v-\perp\bar{P}v=TPu - \perp TPu=0, 
\end{align}
where the last equality holds because directly after the 
application of $T$ the algebraic constraints are satisfied, and 
therefore the projection operator does nothing. Therefore in 
the semi-discrete system if the constraints are satisfied 
initially they remain so, and there is a one-one correspondence 
between solutions of the decomposed and undecomposed systems.

For the fully discrete system we take an explicit polynomial 
time-integrator $Q$ for the undecomposed variables and the 
modified time integrator $\perp Q$ for the decomposed system. 
Now consider the difference between constraint satisfying data
$u^n$ and $v^n=Tu^n$ integrated with the two methods. For brevity 
we subsume the timestep~$\Delta t$ into $P$ and $\bar{P}$. One 
finds that 
\begin{align}
v^{n+1}-Tu^{n+1}&=\perp Q[\bar{P}][v^n] - T Q[P][u^n]\nonumber\\
&= \perp Q[TPS][Tu^n] - T Q[P][u^n]\nonumber\\
&= \perp T Q[P][u^n] - T Q[P][u^n]\nonumber\\
&= T Q[P][u^n] - T Q[P][u^n]=0,
\end{align}
where we have used linearity of the system, polynomiality of 
the time-integrator and the fact that directly after the 
application of $T$ the algebraic constraints are automatically 
satisfied, so the projection operator does nothing. Thus the two 
integration methods are equivalent as desired. Note that in 
these calculations the modified time integrator $\perp Q$ could 
be replaced by $Q$ since the unprojected time-step introduces 
no constraint violation. We have verified these calculations by 
explicitly comparing evolutions of the linearized conformal LaSh 
system with and without constraint projection. We prefer to discuss 
the natural linearization of the non-linear method, which includes 
the projection. In the non-linear case the unprojected timestep 
can introduce algebraic constraint violations. 

\section{Numerical Experiments}\label{sec:numerics}

The LaSh system is implemented inside the \textsc{Lean} 
code~\cite{Sperhake:2006cy} which is based on the
{\sc Cactus} computational toolkit \cite{cactusweb} and the
mesh refinement package {\sc Carpet} \cite{carpetweb,Schnetter:2003rb}.
Initial data is constructed by solving the constraint equations
with the {\sc TwoPunctures} spectral solver provided by 
\cite{Ansorg:2004ds}.

We perform the following set of 
numerical evolutions:
\begin{description}
\item[Robust stability tests] We perform a subset of the 
so-called apples with apples tests \cite{Alcubierre:2003pc}
to demonstrate numerically 
that the evolution system is formally stable with various 
choices of the densitization parameters;

\item[Puncture stability] We evolve a single BH 
with different choices of the densitization parameters to 
establish what restriction is placed on them by insisting on
{\it long-term} stable puncture simulations;

\item[Head-on collisions] We compare BSSN evolutions of the 
head on collision of two BHs with those performed 
with LaSh. We focus on consistency of the extracted physics at 
finite resolution;

\item[Binary black hole inspiral] We compare BSSN evolutions 
of inspiraling BHs (Goddard R1~\cite{Baker:2006yw}) 
with those performed with LaSh. We consider the computational
costs of simulations with the two systems as well as consistency
of the results.
\end{description}

\subsection{Robust stability}

In Sec.~\ref{sec:StabilityAnaDiscrete} we did not succeed in 
demonstrating numerical stability of the LaSh system with the 
puncture gauge using standard discretization. We therefore 
perform robust stability tests following the method in 
Ref.~\cite{Alcubierre:2003pc}. The numerical domain is given by 
$-0.5~<~x~<~0.5$, $-0.06/\Delta~<~y~<~0.06/\Delta$ and 
$-0.06/\Delta~<~z~<~0.06/\Delta$ 
with periodic boundary conditions.
We use three different resolutions $h = 0.02/\Delta$,
where $\Delta_c = 1$, $\Delta_m = 2$ and $\Delta_f = 4$.
The expansion in the $y$- and $z$- direction incorporates 
the three grid points required for fourth order FD stencils.
The initial data are given by small perturbations of the
Minkowski spacetime
\begin{equation}
  \gamma_{ij} = \eta_{ij} + \epsilon_{ij}\,.
\end{equation}
The $\epsilon_{ij}$ are independent random numbers in the range 
$(-10^{-10}/ \Delta, 10^{-10}/ \Delta)$, so that
terms of the order $\mathcal{O}(\epsilon^2)$ are below round-off 
accuracy. This means that the evolution remains in the linear 
regime unless instabilities occur. 

\begin{figure*}
\begin{center}
\begin{tabular}{ccc}
\includegraphics[width=6.4cm]{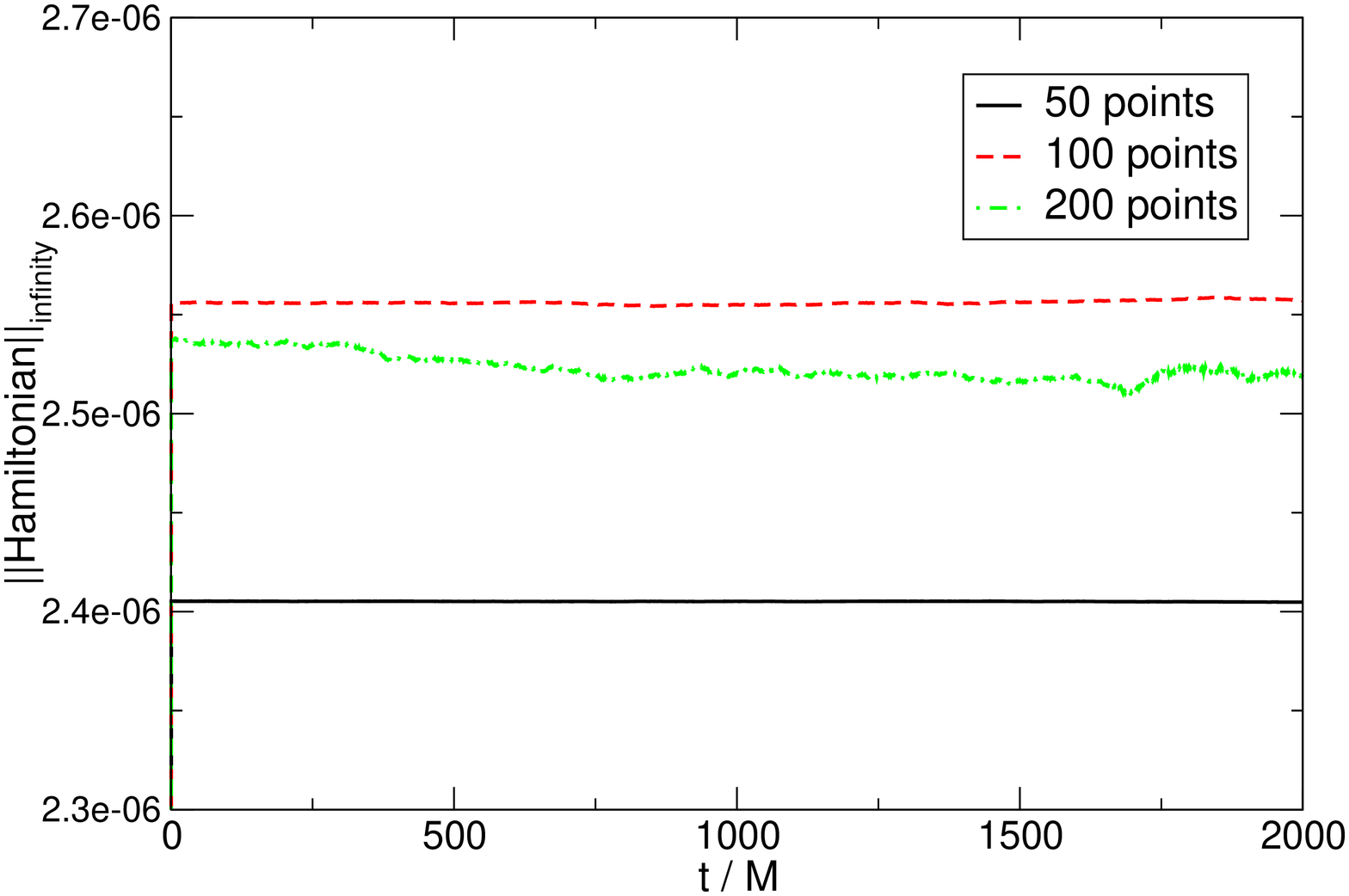} &
\hspace{-0.6cm}
\includegraphics[width=6.4cm]{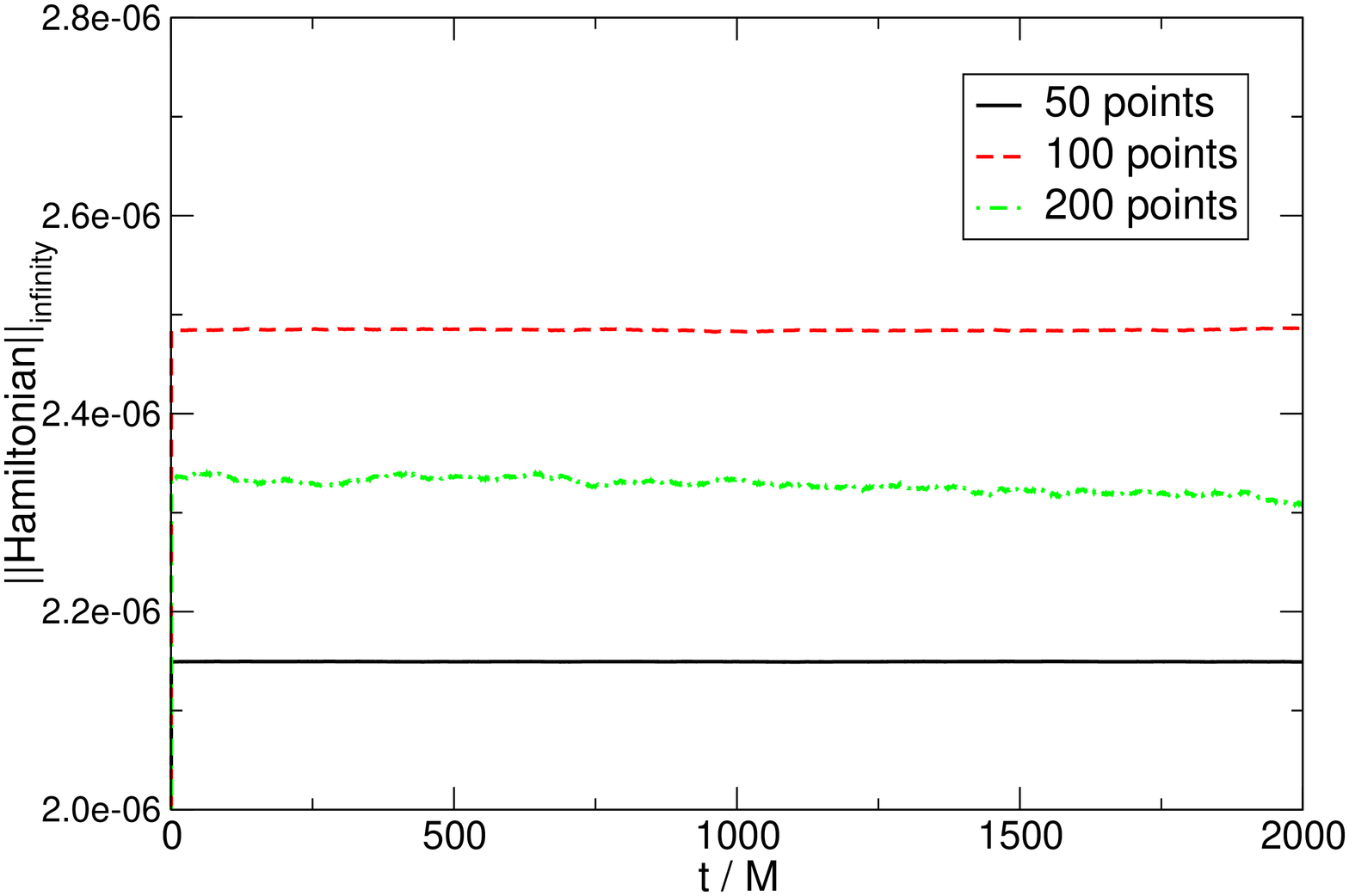} &
\hspace{-0.6cm}
\includegraphics[width=6.4cm]{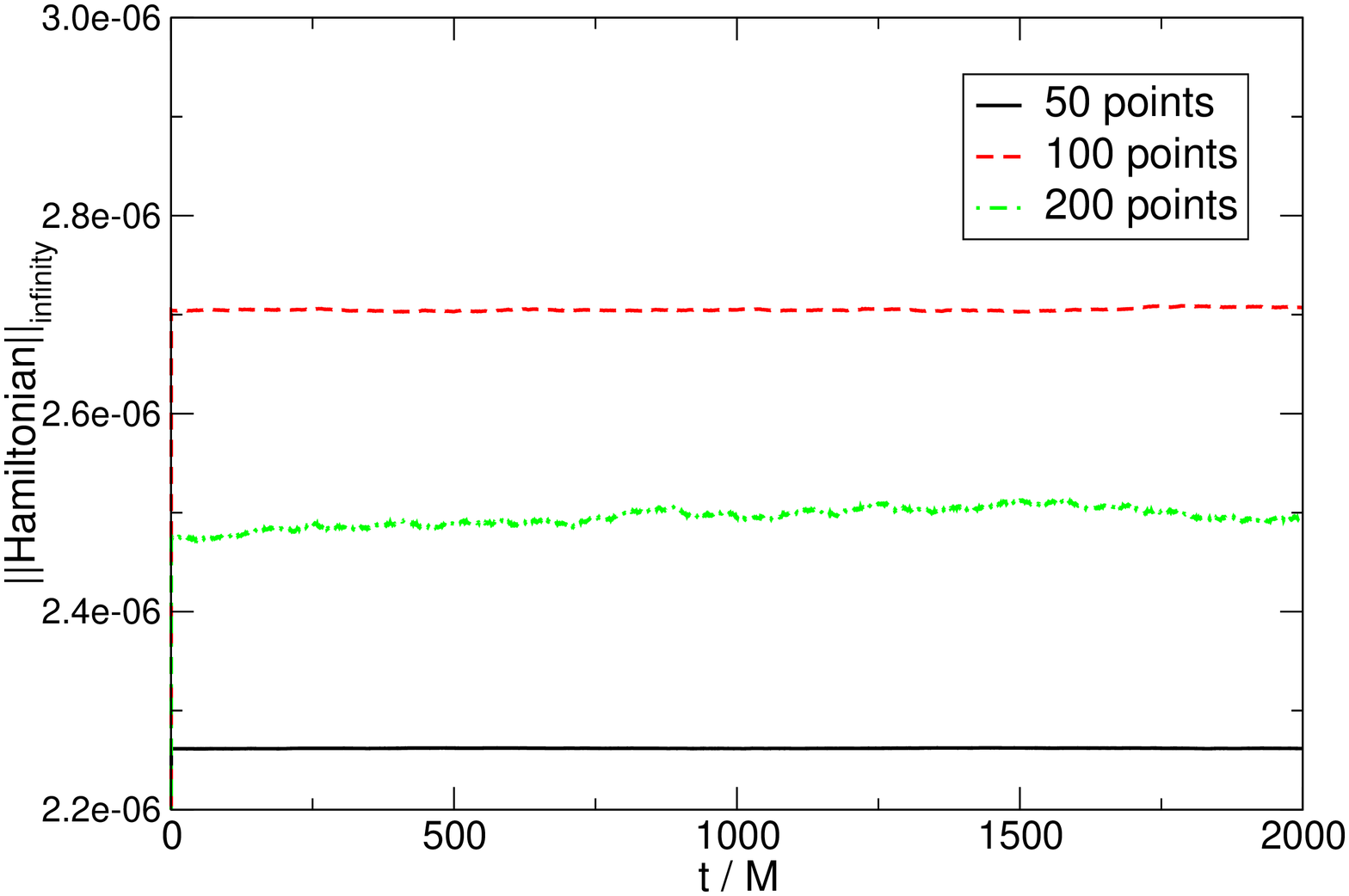} 
\end{tabular}
\end{center}
\caption{\label{fig:statestnqnk} (Color online) 
Apples with apples stability test using a low (50 
grid points, black solid lines), 
medium (100 grid points, red dashed lines)
 and high (200 grid points, green dashed-dotted lines) resolution. 
The tests were performed with the densitization parameters 
$(n_Q,n_K)=\{(0,0),(0.5,-0.5),(-0.5,0.5)\}$ from left to right, 
respectively.
}
\end{figure*}

We monitor the performance of each simulation by
calculating the maximum norm of the Hamiltonian constraint
as a function of time. For this study we focus on three
choices of the densitization parameters $(n_Q,n_K)=\{(0,0),(0.5,-0.5),
(-0.5,0.5)\}$. The results of the robust stability test are plotted 
in Fig.~\ref{fig:statestnqnk}. For all choices of $(n_Q,n_K)$,
including the BSSN scaling $(0,0)$, we obtain stable 
evolutions. 

\subsection{Puncture stability}
\label{sec:vardens}
\begin{figure}[htpb!]
\begin{center}
\includegraphics[width=9cm]{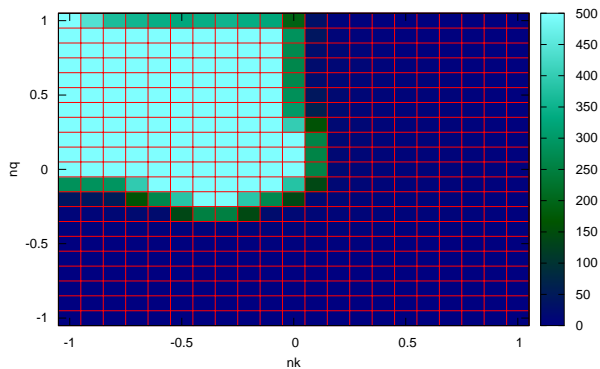}
\caption{\label{fig:ContourParVar} (Color online)
Contourplot of the lifetime $T_l$ as function of the 
densitization parameter $n_Q$ and $n_K$. Areas colored in 
dark blue indicate a short lifetime whereas light blue coloring 
stands for a lifetime of at least $T_l=500M$.
}
\end{center}
\end{figure}

We next perform evolutions of a single puncture, studying a 
wide range of non-trivial densitization parameters.
The hyperbolicity analysis of the continuum LaSh scheme 
presented in Sec.~\ref{sec:StabilityAnaDiscrete} is not 
affected by the choice of densitization parameters provided 
that the algebraic constraints are enforced. In the previous
section we have seen that various choices of the densitization
parameters 
yield evolution systems that are numerically stable. Here we 
demonstrate that those parameters must be chosen more carefully
in order to achieve long-term evolutions of puncture data. 
We evolve a single, non-rotating BH until $t=500M$.
The BH is initially given by two punctures with mass 
parameter $m_{1,2}=0.5M$ located at $z=\pm 10^{-5}M$. Using the notation
of Sec.~II E of Ref.~\cite{Sperhake:2006cy}, the grid 
setup is given in units of $M$ by
\begin{equation}
\{(96, 48, 24, 12, 6, 2, 1, 0.5), 1/32 \}\,.\nonumber
\end{equation}
We vary both $n_Q$ and $n_K$ in the interval $[-1,1]$ in steps 
of $\Delta n = 0.1$. The lifetimes $T_l$ of the simulations are 
determined as functions of the densitization parameters. The 
first occurrence of ``nans" in the right-hand side of 
the densitized lapse $Q$ is used as a measure of the lifetime 
whenever the simulations did not survive for the entire evolution. 
In Fig.~\ref{fig:ContourParVar} we show the results of this
parameter study as a contour plot. In particular,
a single puncture can be evolved for at
least $t=500~M$ using the LaSh system with parameters indicated
by the light blue area in the figure.
Negative values of the 
lapse densitization parameter $n_Q < -0.3$ combined with 
positive values of the curvature densitization parameter 
$n_K > 0$ 
let the simulations crash after a short time.
In contrast, long term stable evolutions are obtained
for the parameter range $n_Q\in [-0.3,0.9]$,
$n_K \in [-1,0]$, including the BSSN scaling $n_Q=n_K=0$.

We can partially understand this behaviour
by considering single puncture initial data and their influence on 
the evolution equations~(\ref{LaSh-g}-\ref{LaSh-Gamma}).
On the initial slice the densitized lapse $Q$ is given by
\begin{align}
Q_0 = \chi^{\frac{3n_Q}{2}}\alpha_0 = \chi^{\frac{3}{2}(n_Q+1/3)},
\end{align}
corresponding to a {\it pre collapsed} lapse $\alpha$.
Since $\chi$ vanishes at the puncture we require $n_Q > -\frac{1}{3}$ to 
obtain a regular densitized lapse $Q_0$ on the initial timeslice,
in agreement with the findings of our parameter study;
simulations with $n_Q < -0.3$ crash immediately.
Next consider the evolution 
equations on the initial timeslice. For our initial data they reduce to
\begin{align}
\label{LaSh-g-init}
\partial_t\tilde{\gamma}_{ij} = 0, &\qquad 
\partial_t \chi = 0\,,\\
\label{LaSh-A-init}
\partial_t \tilde{A}^i{}_j = &
\chi^{-3n_K/2}[D^iD_j\alpha+\alpha R^i{}_j]^{TF}\,,\\
\label{LaSh-K-init}
\partial_t\tilde{K} = & - \chi^{-3n_K/2}D^iD_i\alpha\,,\\
\label{LaSh-Gamma-init}
\partial_t \tilde{\Gamma}^i= & 0\,.
\end{align}
Insisting on initially regular evolved variables at the puncture, 
Eqs.~(\ref{LaSh-g-init}-\ref{LaSh-Gamma-init}) require $n_K\leq 0$,
also in agreement with our study;
numerical experiments violating this condition 
immediately fail.

\begin{figure*}
\begin{center}
\begin{tabular}{cc}
\includegraphics[width=8.0cm]{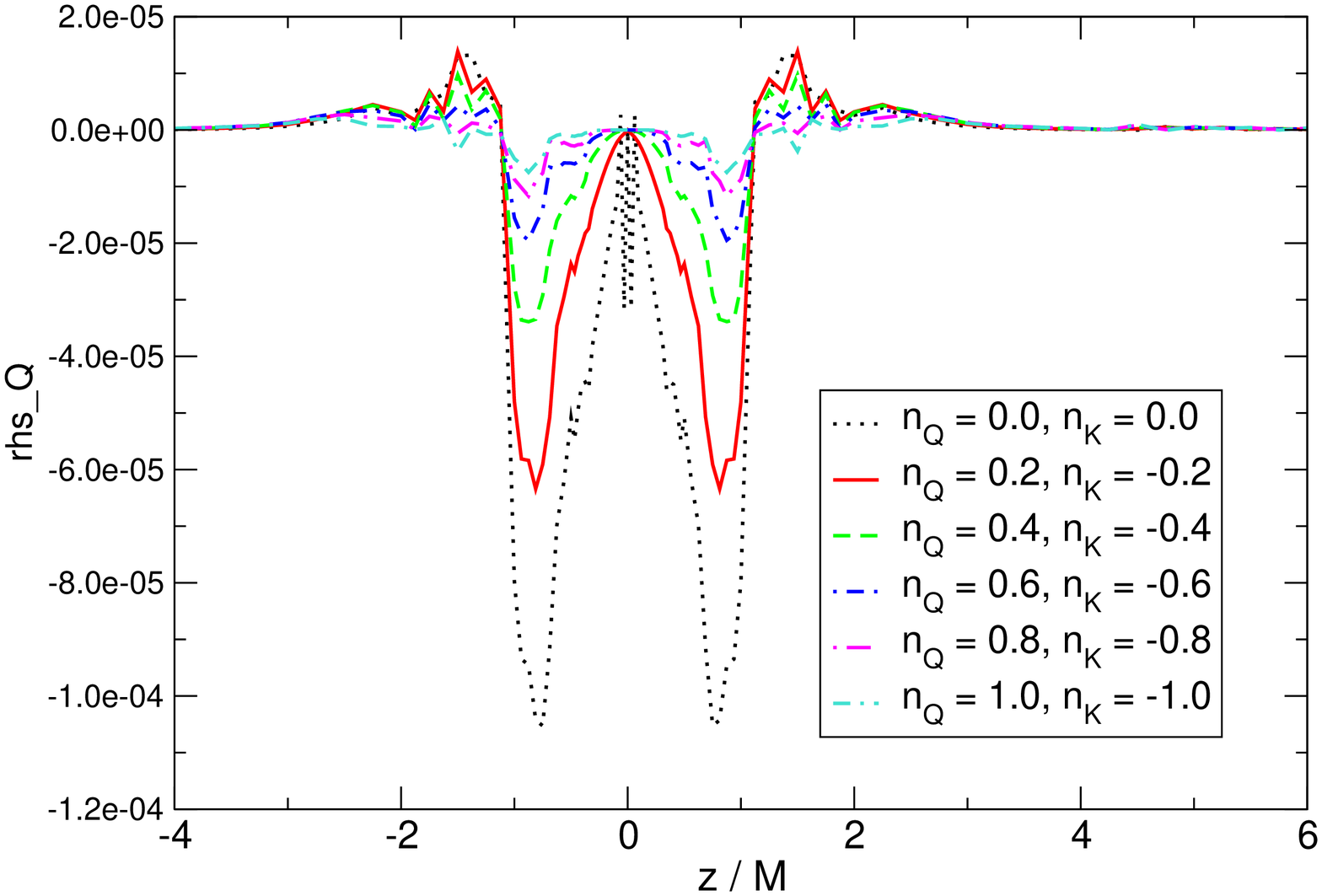} &
\includegraphics[width=8.0cm]{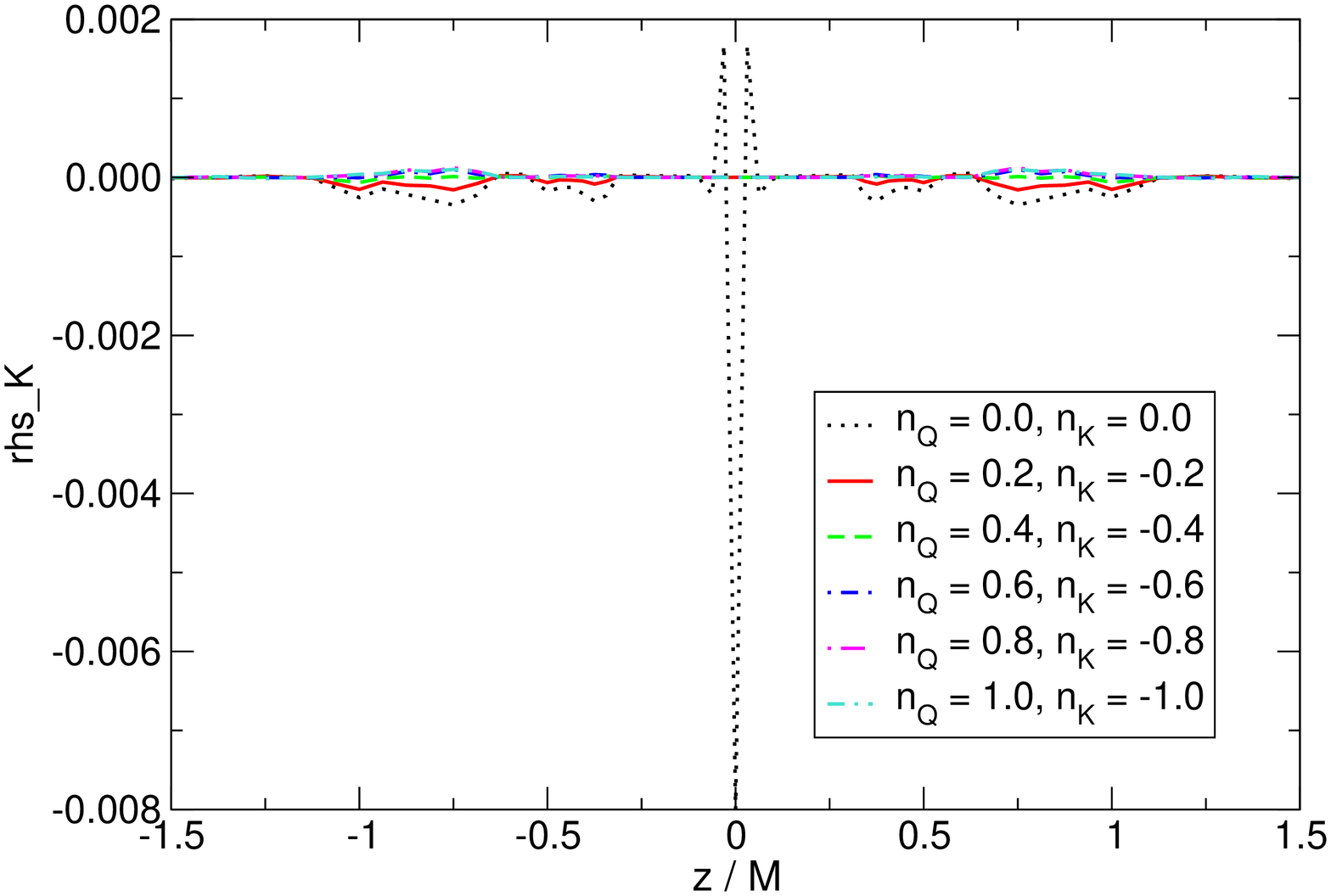}
\end{tabular}
\end{center}\caption{\label{fig:Rhs100MParVar} (Color online)
  Right hand sides of the densitized lapse $Q$ (left panel) and 
  of the trace of the extrinsic curvature $\tilde{K}$ (right panel) 
  after an evolution time of $t=100M$. We take parameter pairs 
$(n_Q,n_K)=\{(0,0),(0.2,-0.2),(0.4,-0.4),(0.6,-0.6),(0.8,-0.8),(1,-1)\}$.
}
\end{figure*}

For further illustration
we plot in Fig.~\ref{fig:Rhs100MParVar}
the time derivatives of the densitized lapse $Q$ 
and the trace of the extrinsic curvature $\tilde{K}$
after an evolution time of $t=100~M$. As we simoultaneously
increase $n_Q$ and decrease $n_K$, we obtain smoother profiles.
Note that the BSSN case $n_Q=0=n_K$ produces the steepest gradients
in this comparison. A systematic study of the
exceptionally benign behaviour of a non-trivial densitization
on the accuracy of 10-15 orbit simulations, especially
of spinning, precessing binaries, is beyond the scope of this paper.
Our results may, however, point at fertile ground for future
research of the LaSh system.

\subsection{Head-On Collisions}

\begin{figure}
\begin{center}
\includegraphics[width=8.0cm]{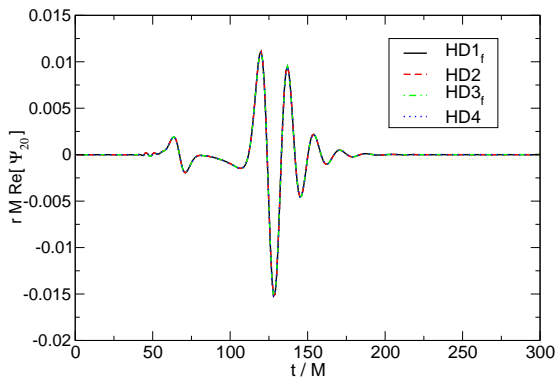} 
\end{center}
\caption{\label{fig:HDwaveforms}(Color online)
Real part of $r_{\rm ex}M\Psi_{20}$,
the dimensionless Newman-Penrose scalar, 
where $r_{ex} = 60M$,
for model $HD1_f$ (black solid line), $HD2$ (red dashed line),
$HD3_f$ (green dashed-dotted line) and $HD4$ (blue dotted line).
}
\end{figure}

\begin{table*}
\begin{tabular}{|c|c|c|c|c|c|}
\hline
\parbox[]{1cm}{\centering Run}          & 
\parbox[]{3.5cm}{\centering Grid Setup}   & 
\parbox[]{1.0cm}{\centering $d/M$}        &
\parbox[]{1.0cm}{\centering $n_Q$}  &
\parbox[]{1.0cm}{\centering $n_K$}  &
\parbox[]{1.7cm}{\centering $10^4 E_{rad}/M$ }
\\ \hline
$HD1_c$ & $\{(256,128,72,32,16)\times (4,2,1),~h=1/40\}$ & $10.24$ & $0.0$ & $0.0$ & $5.51$  \\
$HD1_m$ & $\{(256,128,72,32,16)\times (4,2,1),~h=1/44\}$ & $10.24$ & $0.0$ & $0.0$ & $5.52$ \\
$HD1_f$ & $\{(256,128,72,32,16)\times (4,2,1),~h=1/48\}$ & $10.24$ & $0.0$ & $0.0$ & $5.53$  \\
\hline
$HD2$ & $\{(256,128,72,32,16)\times (4,2,1),~h=1/48\}$ & $10.24$ & $0.2$ & $-0.2$ & $5.53$  \\
\hline
$HD3_c$ & $\{(256,128,72,32,16)\times (4,2,1),~h=1/40\}$ & $10.24$ & $0.4$ & $-0.4$ & $5.51$  \\
$HD3_m$ & $\{(256,128,72,32,16)\times (4,2,1),~h=1/44\}$ & $10.24$ & $0.4$ & $-0.4$ & $5.52$  \\
$HD3_f$ & $\{(256,128,72,32,16)\times (4,2,1),~h=1/48\}$ & $10.24$ & $0.4$ & $-0.4$ & $5.53$  \\
\hline
$HD4$ & $\{(256,128,72,32,16)\times (4,2,1),~h=1/48\}$ & $10.24$ & $0.6$ & $-0.6$ & $5.53$  \\
\hline
\end{tabular}
\caption{\label{tab:headonruns} 
  Grid structure and physical initial parameters of the simulations of a
  head-on collsion of an equal mass BH binary.
  The grid setup is given in terms of the radii of the individual 
  refinement levels as well as the resolution near the punctures $h$ 
  (see Sec.~II E in \cite{Sperhake:2006cy} for details). 
  The table further shows the initial coordinate separation $d/M$ of the two 
  punctures.
  $E_{rad}/M$ is the fraction of the total BH mass that 
  is radiated as gravitational waves. All parameters are given in units 
  of the total BH mass $M=M_1+M_2$.
}
\end{table*}
In this section we study in depth the stability properties
of numerical simulations
of equal-mass head-on collisions
performed with the LaSh system.
For this purpose we evolve model $BL2$ of Table~II in \cite{Sperhake:2006cy},
i.e. two non-spinning holes with irreducible mass $M_{{\rm irr},i} = 0.5~M$
starting from rest at $z_{1,2} = \pm 5.12~M$.
The computational grid consists of a set of 
nested refinement levels given in units of $M$ by
\begin{align}
  \{(256,128,72,32,16)\times (4,2,1),~h\}\,,
  \nonumber
\end{align}
where we have usually chosen $h = M / 48$, unless denoted otherwise.
We consider the densitization parameters
$(n_Q, n_K)~=~\{(0,0), (0.2, -0.2), (0.4, -0.4), (0.6,-0.6)\}$,
denoted as models $HD1$ - $HD4$ in Table~\ref{tab:headonruns}.
For models $HD1$ - $HD3$ we have chosen the $\Gamma$-driver shift
conditions~\eqref{eqn:LaShshiftch2} with 
$(\mu_S,\xi_1,\xi_2,\eta)=(1,0,0,1)$, whereas 
in case of model $HD4$ the $\Gamma$-driver shift 
conditions~\eqref{eqn:LaShshiftch2} have been taken with 
$(\mu_s,\xi_1,\xi_2,\eta)=(3/4,1,1,1)$ 
\cite{Brugmann:2008zz}.
Information about gravitational waves emitted during the plunge has been obtained 
by the Newman-Penrose scalar $\Psi_4$. In Fig.~\ref{fig:HDwaveforms} we present
the real part of the
dominant mode $\Psi_{20}$, rescaled by the extraction radius $r_{ex} = 60M$,
for models $HD1$-$HD4$.
Note, that the imaginary part of $\Psi_4$ vanishes due to symmetry.
We find that the waveforms generated by the different models agree well.
We study the convergence of models $HD1$ and $HD3$ by using
three different resolutions $h_c=M/40$, 
$h_m = M/44$ and $h_f = M/48$ referred to as {\it coarse}, 
{\it medium} and {\it high} resolution.
\begin{figure*}
\begin{center}
\begin{tabular}{cc}
\includegraphics[width=8.0cm]{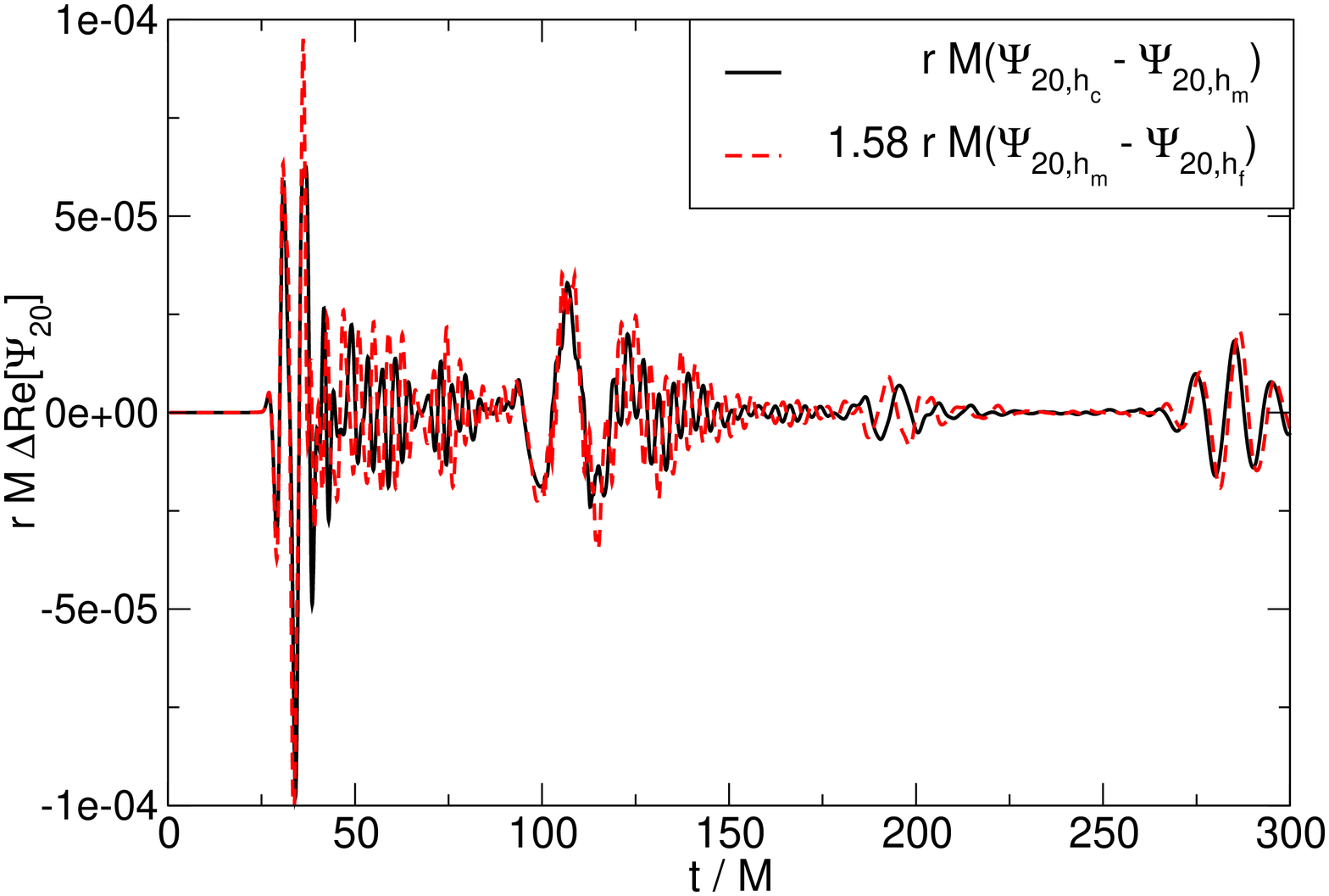} &
\includegraphics[width=8.0cm]{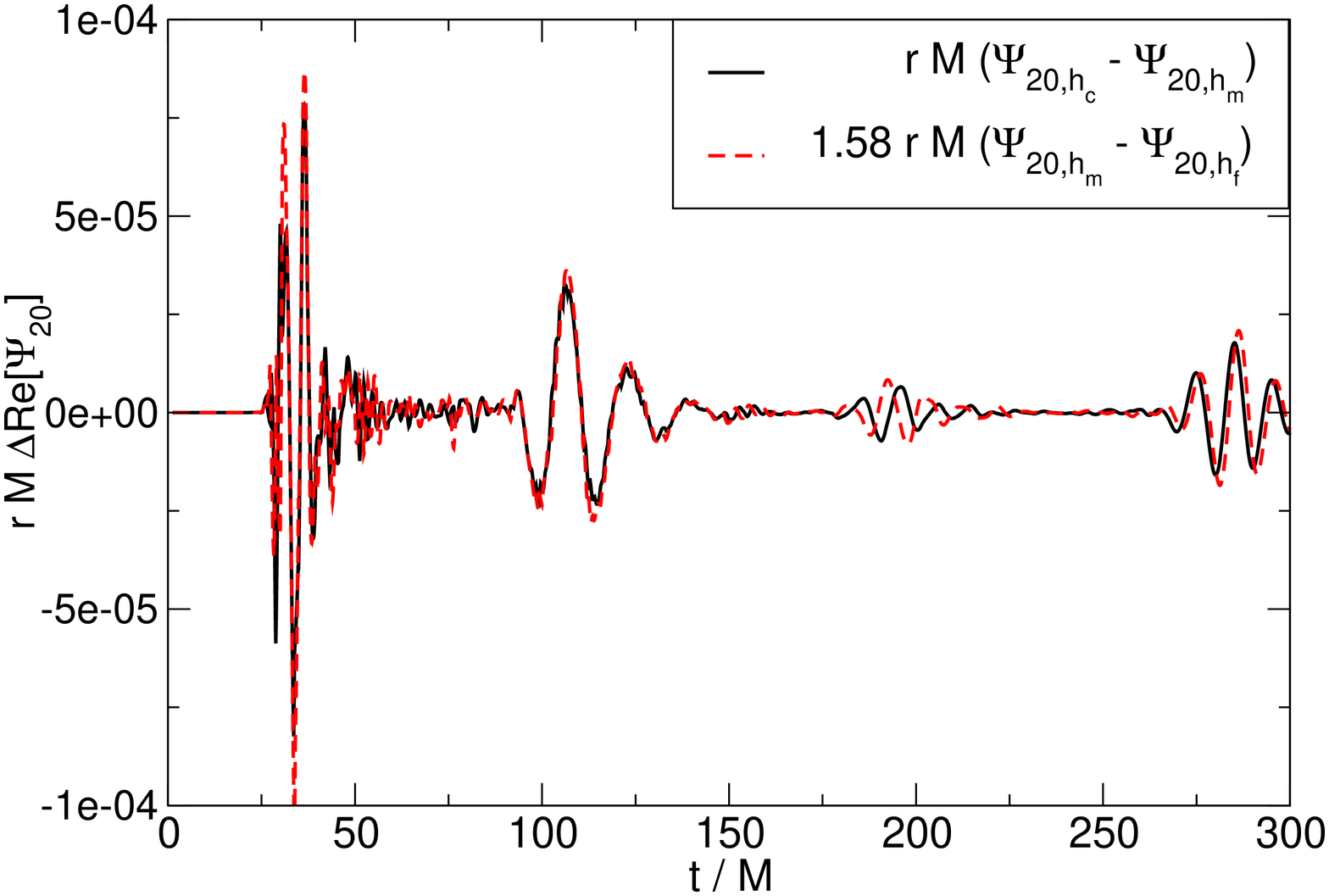}
\end{tabular}
\end{center}
\caption{\label{fig:HDconvergence}(Color online)
Convergence analysis of the real part of $\Psi_{20}$ of the Newman Penrose 
scalar $\Psi_4$, re-scaled by the extraction radius $r_{ex} = 40M$, for models $HD1$ (left panel) and $HD3$ (right panel)
in Table~\ref{tab:headonruns}. 
We show the difference between the low and medium resolution (black solid line) and
the medium and high resolution (red dashed line).
The latter has been amplified by a factor of $Q=1.58$ expected for fourth order convergence 
}
\end{figure*}
The differences of the $\ell=2$, $m=0$ mode of the resulting gravitational
radiation are displayed in Fig.~\ref{fig:HDconvergence} and demonstrate
overall fourth order convergence for both models.
We estimate the discretization error at high resolution
in the waveforms $\Psi_{20}$ to be $0.4\%$, similar to
the error reported in \cite{Sperhake:2006cy} for the corresponding
BSSN evolutions.

The amount of energy that is radiated throughout the head-on collision
computed from, e.g., Eq.~(22) in Ref.~\cite{Witek:2010qc} 
(see also \cite{Alcubierre:2008})
is $E_{rad}/M = 0.0553\%$ for models $HD1_f$, $HD2$, $HD3_f$ and $HD4$,
again in excellent agreement with Ref.~\cite{Sperhake:2006cy}.
We estimate the discretization error in the radiated energy to 
be $0.4\%$ and the error due to finite extraction
radius to be $1.6\%$.

As for single BH evolutions, we observe smoother time derivatives
of $Q$ and $\tilde{K}$ for non-vanishing choices of $n_Q$ and $n_K$.
\begin{figure*}
\begin{center}
\begin{tabular}{cc}
\includegraphics[width=8.0cm]{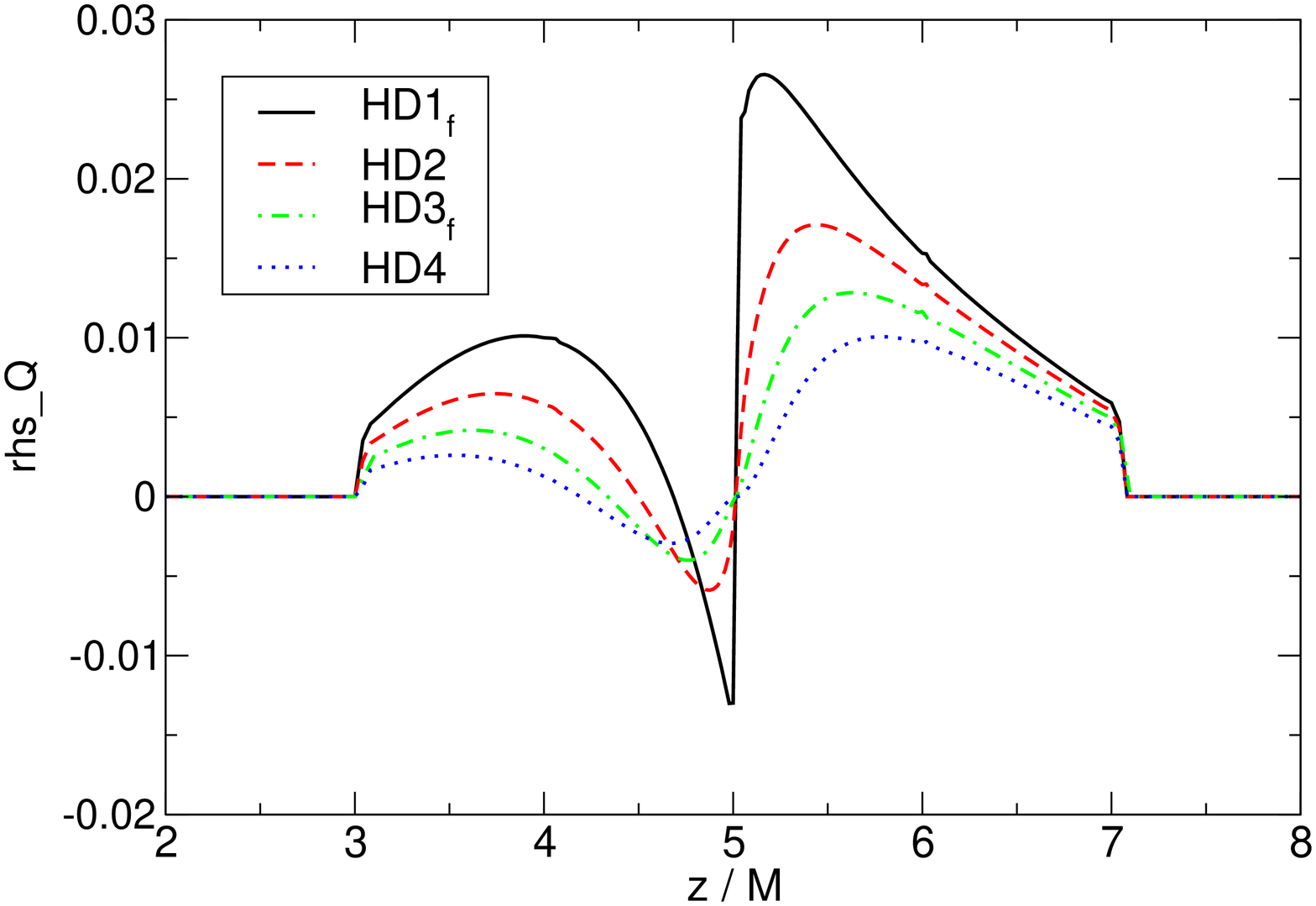} &
\includegraphics[width=8.0cm]{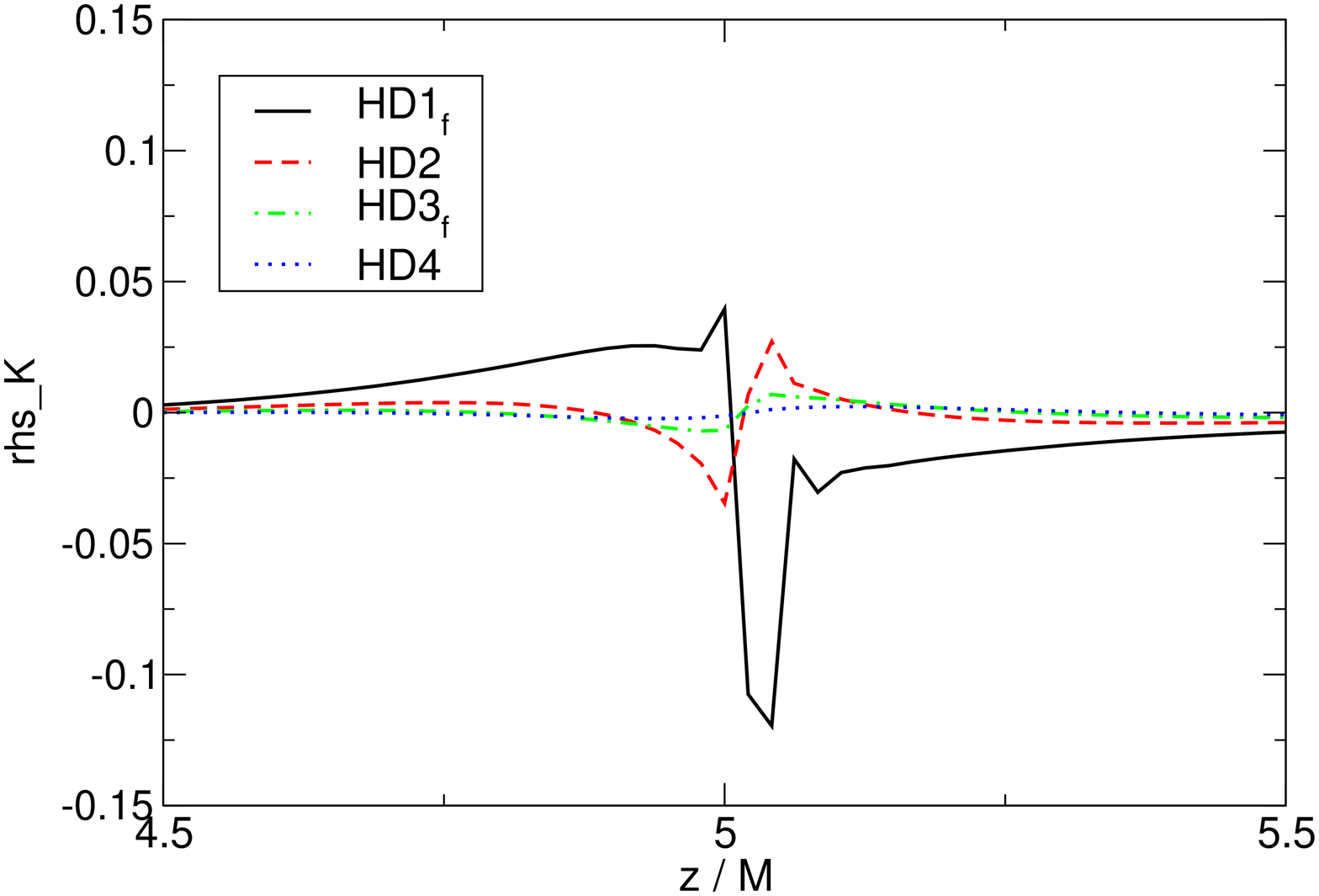}
\end{tabular}
\end{center}
\caption{\label{fig:HD_rhs_t10}
Right hand sides of the densitized 
lapse $Q$ (left panel) and of the trace of the extrinsic curvature $\tilde{K}$
(right panel) after an evolution time of $t=10M$ for models $HD1_f$ (solid line),
$HD2$ (dashed line), $HD3_f$ (dashed-dotted line) and $HD4$ (dotted line).
}
\end{figure*}
We illustrate this behaviour in Fig.~\ref{fig:HD_rhs_t10}
which shows the time derivatives along the $z$ axis obtained
for different values of $n_Q$ and $n_K$ at $t=10~M$.

\subsection{Inspiraling Black-Holes}

\begin{figure*}
\begin{center}
\begin{tabular}{cc}
\includegraphics[width=8.0cm]{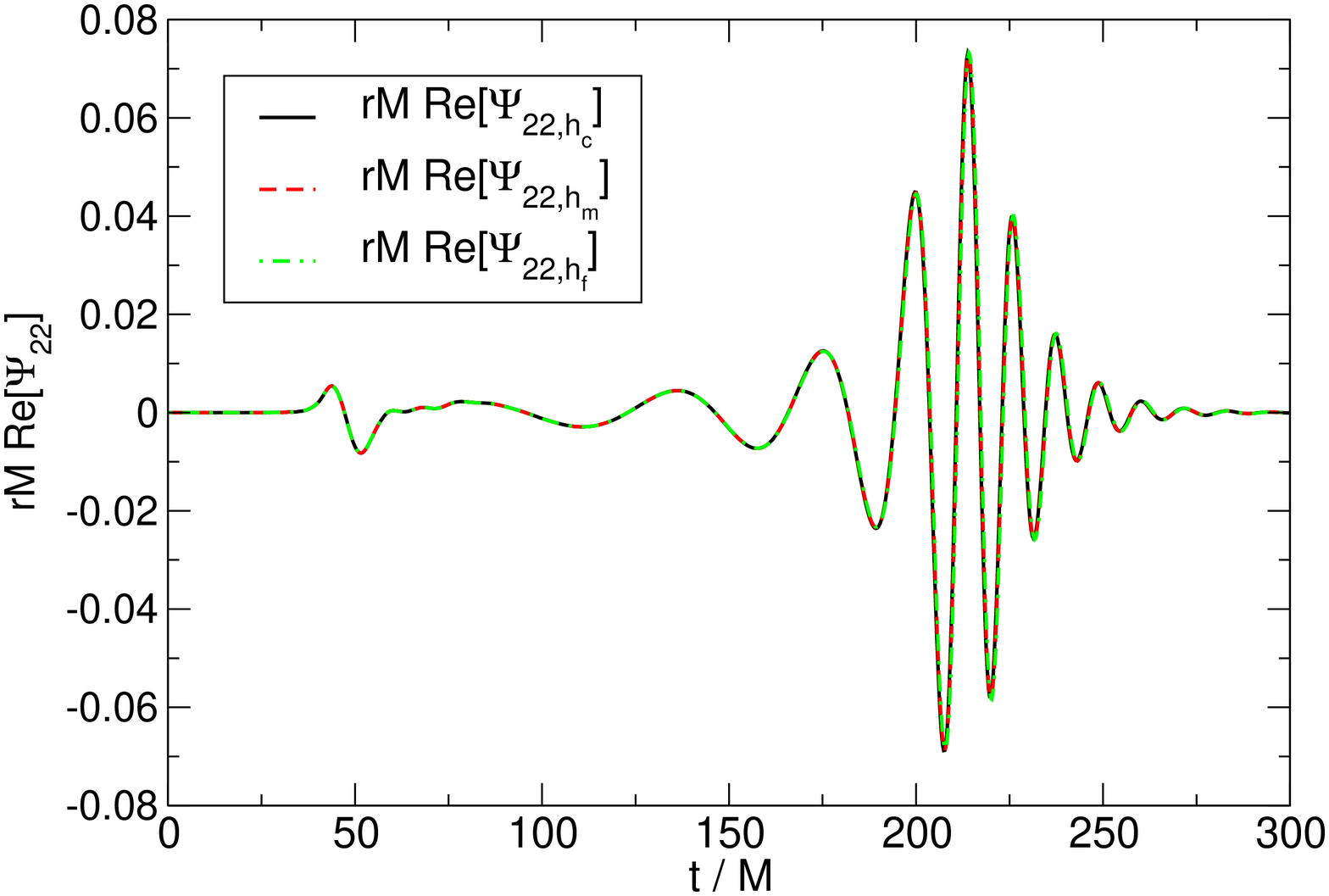} &
\includegraphics[width=8.0cm]{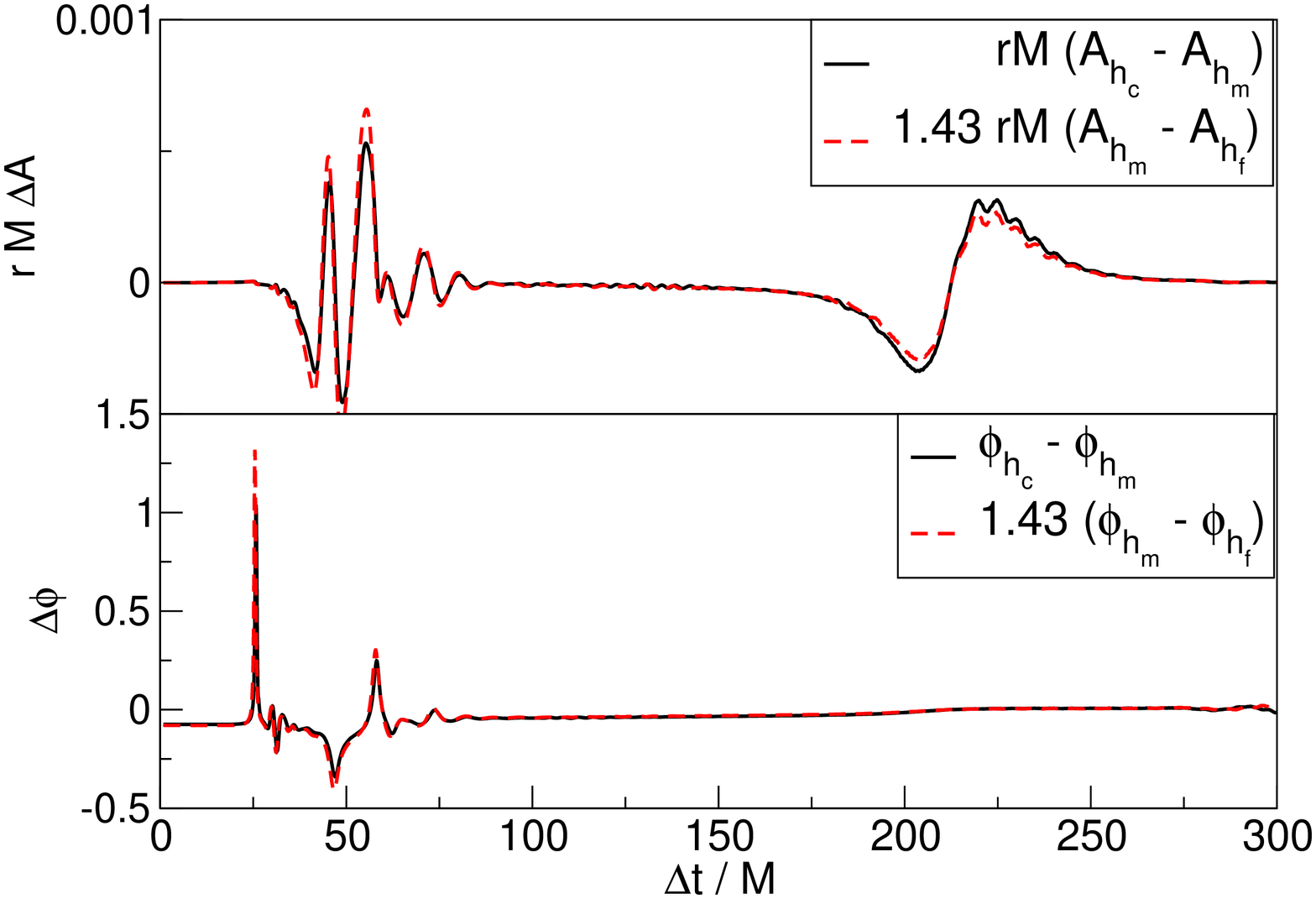} 
\end{tabular}
\end{center}\caption{\label{fig:InspiralConvergence} (Color online)
Left:
  Real part of the dominant $\ell = 2$, $m=2$ mode of the dimensionless
  Newman-Penrose scalar $r M \Re(\Psi_4)$, where the extraction radius is
  $r_{ex}=40M$. The waveforms are shown for models $LaSh_c$ 
  (black solid line), $LaSh_m$ (red dashed line) and $LaSh_f$ 
  (green dash-dotted line). 
Right:
  Convergence analysis of the Amplitude (upper panel) and phase (bottom panel)
  of the dominant $\ell = 2$, $m=2$ mode of the Newman Penrose scalar $\Psi_4$.
  We show the differences between the coarse and medium resolution (black solid line) 
  and medium and high resolution (red dashed line). 
  The latter difference has been amplified by $Q_4 = 1.43$, indicating  
  fourth order convergence.
}
\end{figure*}

\begin{table*}
\begin{tabular}{|c|c|c|c|c|c|}
\hline
\parbox[]{1cm}{\centering Run}                  & 
\parbox[]{2cm}{\centering Evolution Scheme}     & 
\parbox[]{3.5cm}{\centering Grid Setup}         & 
\parbox[]{1.0cm}{\centering $n_Q$}              &
\parbox[]{1.0cm}{\centering $n_K$}              &
\parbox[]{1.7cm}{\centering $10^2 E_{rad}/M$ }
\\ \hline
$BSSN$ & BSSN & $\{(256,128,64,24,12,6)\times (1.5,0.75),~1/48\} $ & - & - & -  \\
\hline
$LaSh_0$ & LaSh & $\{(256,128,64,24,12,6)\times (1.5,0.75),~1/48\}$ & $0.4$ & $-0.4$ &  -  \\
$LaSh_c$ & LaSh & $\{(256,128,64,24,12,6)\times (1.5,0.75),~1/52\}$ & $0.4$ & $-0.4$ & $3.69$  \\
$LaSh_m$ & LaSh & $\{(256,128,64,24,12,6)\times (1.5,0.75),~1/56\}$ & $0.4$ & $-0.4$ & $3.68$  \\
$LaSh_f$ & LaSh & $\{(256,128,64,24,12,6)\times (1.5,0.75),~1/60\}$ & $0.4$ & $-0.4$ & $3.67$  \\
\hline
\end{tabular}
\caption{\label{tab:inspiralruns}
  Grid structure, evolution system and initial parameters of the simulations
  of quasi-circular inspirals. The grid setup is given in terms of the 
  radii in units of $M$
  of the individual refinement levels as well as the resolution 
  near the punctures $h$ (see Sec.~II E in \cite{Sperhake:2006cy} 
  for details). In case of the LaSh scheme we also 
  specify the densitization parameters $n_Q$ and $n_K$.
  The final column lists the radiated energy $E_{\rm rad}$
  extracted at $r_{\rm ex} = 60~M$ for models $LaSh_c$-$LaSh_f$.
  Models $BSSN$ and $LaSh_0$ have only been run until $t=50M$ in order 
  to compare their computational cost.
}
\end{table*}
In this section we will demonstrate how BBHs
can be evolved successfully using the LaSh formulation of the
3+1 Einstein equation in combination with the moving puncture
approach.
For this purpose we consider the initial configuration labeled
R1 in Table~I of Ref.~\cite{Baker:2006yw}. 
This configuration
represents a non-spinning, equal-mass binary with a total ADM mass
of $M=0.9957$ in code units. The bare-mass parameters
are $m_{1,2} = 0.483$ 
and the BHs start at position
$x_{1,2} = \pm 3.257$ with linear momentum
$P_{1,2} = \pm 0.133$ in the $y$-direction.
The specifications of the grid setup, 
in the notation of Sec.~II E of Ref.~\cite{Sperhake:2006cy},
are given in Table~\ref{tab:inspiralruns}. 
For this model we have used
the $\Gamma$-driver shift condition~\eqref{eqn:LaShshiftch2} 
with $(\mu_s,\xi_1,\xi_2,\eta)=(3/4,1,1,1)$ as suggested
in Ref.~\cite{Brugmann:2008zz}.
As before, we study the convergence properties by performing simulations
of model $LaSh$ with
resolutions $h_c = 1/52$, 
$h_m = 1/56$ and $h_f = 1/60$.
In the left panel of Fig.~\ref{fig:InspiralConvergence} we present the 
real part of the $\ell = 2$, $m=2$ mode of $\Psi_4$, extracted at
$r_{ex} = 40M$, obtained by models $LaSh_c$, $LaSh_m$ and $LaSh_c$.
The right panel of Fig.~\ref{fig:InspiralConvergence} shows the 
differences between the coarse and medium and medium and high resolutions
of the amplitude (upper panel) and phase (bottom panel).
The latter differences have been rescaled by the factor $Q_4 = 1.43$ 
corresponding to fourth order convergence.
The resulting discretization error in amplitude and phase are
$\Delta A / A \leq 1\%$ and $ \Delta \phi \leq 0.1~{\rm rad}$.
 
The energy
radiated in gravitational waves is
$E_{rad}/M = 3.67 \pm 0.13 \%$ for 
the high resolution run
$LaSh_f$ in Table~\ref{tab:inspiralruns} which is in good agreement
with the BSSN results of Ref.~\cite{Sperhake:2006cy}.
 
\begin{table}[htpb!]
\begin{center}
\begin{tabular}{|c|c|c|c|}
\hline
\parbox[]{1.0 cm}{\centering Run} &
\parbox[]{1.5 cm}{\centering mem. [GByte]} &
\parbox[]{2.0 cm}{\centering $t_r$ [CPUhours]}&
\parbox[]{1.5 cm}{\centering $\bar{v}$ [M/hour]}\\
\hline
\hline
$BSSN$    & $55$  & $290$ & $4.2$ \\
\hline
$LaSh_0$  & $70$  & $430$ & $2.9$ \\
\hline
$modLaSh$ & $55$  & $335$ & $3.7$ \\
\hline
\end{tabular}
\caption{\label{tab:performance} 
  The required memory $mem.$, the total runtime $t_r$ in CPUhours 
  and the average speed $\bar{v}$ in units of physical time $M$ per 
  real time hour  of the test simulations using the BSSN (model $BSSN$ 
  in Table~\ref{tab:inspiralruns}), the original LaSh (model $LaSh_0$ 
  in Table~\ref{tab:inspiralruns}) and the modified LaSh scheme.
  The simulations have been run for $t=50M$ using $24$ processors.
}
\end{center}
\end{table}

Finally, we compare the computational performance of 
both, the BSSN and LaSh evolution scheme.
For this purpose we have evolved models $LaSh_0$ and $BSSN$ until $t=50M$
on the Magerit cluster \cite{magerit2010} in Madrid which is part of the 
Spanish Supercomputing Network~\cite{BSC-RES2010}.
Magerit uses PowerPC-970FC processors running at $2.2GHz$. 
The required memory,
runtime and average speed obtained for $24$ processors are 
shown in Table~\ref{tab:performance}.
The original LaSh system requires about $30~\%$ 
more memory than the BSSN system and is about a
factor $1.4$ slower. The overhead of the LaSh system is not
unexpected. First, the LaSh system involves a larger number
of grid functions; the tracefree part of the extrinsic
curvature $\hat{A}^i{}_j$ is not symmetric and thus requires
$9$ independent components instead of $6$ for the
BSSN variable $\tilde{A}_{ij}$. Second, the densitization
of variables requires extra variables
and involved more complicated expressions on
the right hand sides of the corresponding evolution equations.
These effects can be partly eliminated, however, without loosing the
appealing properties of the LaSh system. For this purpose, we have
tested a modified version of the original LaSh system,
denoted as $modLaSh$ in Table~\ref{tab:performance}. Here
we evolve $\tilde{A}_{ij}$ instead of the trace-free part of 
the extrinsic curvature with mixed indices $\tilde{A}^i_j$.
As expected, this modification equals the BSSN system in memory
requirements and also significantly reduces the computational
costs relative to the original LaSh system. At the same time,
however, $modLaSh$ preserves the flexibility that has enabled us
to obtain smoother behaviour of the
variables close to the puncture as compared with the BSSN scheme.

\section{Conclusions}\label{sec:conclusion}

Motivated by a desire to better understand which are the 
important ingredients of the moving puncture method, we have 
studied the LaSh formulation of the Einstein equations. 
Provided that the algebraic constraints of the system are 
imposed the formulation is equivalent to BSSN. Therefore we 
have investigated how the choice of evolved variables effects 
the success of numerical simulations of puncture initial data.
The change of variable is parametrized by the densitization 
parameters $(n_Q,n_K)$.
 
We started by demonstrating that LaSh is formally numerically 
stable when linearized around flat space for arbitrary 
densitization parameters, with fixed shift and densitized 
lapse. A special case of this calculation is the numerical 
stability of BSSN. We attempted to show numerical stability 
of the system coupled to the puncture gauge, but find that 
the required calculations are too complicated even for computer 
algebra unless we move away from the standard discretization. 

We performed four types of numerical tests.
The first class of tests includes robust stability test,
specifically the so-called apples with apples tests.
We find that the LaSh formulation is numerically stable 
for various choices of the densitization parameters. 
Next, we found that long term stable evolutions of single BH spacetimes
require a more careful choice of the densitization parameters.
It is interesting to note that the parameter choice corresponding
to the BSSN system is located near the edge of the permissible
range. 
Moreover, we have identified
parameter choices which result in smoother profiles of the
time derivatives of the evolution variables
near the puncture as compared with the BSSN case.
It will be interesting to investigate the impact of this behaviour
on the accuracy of inspiral simulations lasting $10-15$ orbits.
While such a study is beyond the scope of this paper, it may
provide fertile ground for direct application of the results presented
in this work.
Furthermore, preliminary tests of higher dimensional BH spacetime
evolutions indicate that the generalized BSSN formulation helps overcoming 
stability problems that have been encountered in $D\geq6$ \cite{Zilhao:2010sr}.
A more detailed analysis of this application will be presented elsewhere \cite{HD_LaSh_inprep}.

We have further evolved head-on collisions as well
as quasi-circular inspirals of binary BHs. For both cases, we have
achieved long-term stable evolutions for a wide range of non-trivial 
parameter choices $(n_Q,n_K)$. The evolutions produce convergent 
waveforms consistent with the BSSN results and comparable accuracy.
As mentioned above, we plan to compare the accuracy of both systems
for more demanding inspiral simulations in future work.
In any case, the binary simulations confirm the above finding that
non vanishing values of $n_Q$ and $n_K$
facilitate evolutions with smoother profiles of the evolution
variables in the neighborhood
of each puncture.

In summary, our results highlight the importance of the choice of variables
for numerical calculations aside from any continuum PDE 
considerations. This opens up the possibility of significantly
reducing errors in simulations of astrophysical binaries
with large spins or mass ratios and also overcome
stability issues reported for higher dimensional BH
simulations \cite{Zilhao:2010sr}.

\begin{acknowledgments}
The authors are grateful to Emanuele Berti, Bernd Br\"ugmann, 
Ian Hinder, Ronny Richter and Olivier Sarbach for helpful comments on 
the manuscript. H.W.'s work was partly supported by DFG grant 
SFB/ Transregio7 and by Funda\c c\~ao para a Ci\^encia e Tecnologia 
(FCT) - Portugal through grant SFRH/BD/46061/2008 and 
through projects PTDC/FIS/098032/2008, PTDC/FIS/098025/2008 and CERN/FP/116341/2010.
D.H. was supported partly by DFG grant SFB/Transregio7.
U.S. acknowledges support from the Ram\'on y Cajal Programme of
the Spanish Ministry of Science and Innovation (MICINN) as well
as FCT -- Portugal through project PTDC/FIS/098025/2008,
the Sherman Fairchild Foundation to Caltech, by
NSF grants PHY-0601459, PHY-0652995, PHY-1057238,
by loni\_numrel05,
allocations through the TeraGrid Advanced Support
Program under grants PHY-090003
and AST-100021 on NICS' Kraken and SDSC's Trestless clusters and an allocation
by the Centro de Supercomputaci\'on de Galicia (CESGA) under Project Nos. ICTS-CESGA 120 and ICTS-CESGA 175 on Finis Terae.
This work was also supported by NSF Grant PHY-0900735
and by the {\it DyBHo--256667} ERC Starting Grant.
Computations were partly performed at the LRZ Munich, Milipeia in Coimbra
and Magerit in Madrid.
The authors thankfully acknowledge the computer resources, technical 
expertise and assistance provided by the Barcelona Supercomputing 
Centre---Centro Nacional de Supercomputaci\'on.
\end{acknowledgments}

\bibliographystyle{h-physrev4}
\bibliography{LaSh_v2}

\begin{thebibliography}{10}

\bibitem{McClintock:2003gx}
J.~E. McClintock and R.~A. Remillard,
\newblock astro-ph/0306213.

\bibitem{Rees1984}
M.~Rees,
\newblock Ann. Rev. Astron. Astrophys. {\bf 22}, 471 (1984).

\bibitem{Ferrarese:2004qr}
L.~Ferrarese and H.~Ford,
\newblock Space Sci. Rev. {\bf 116}, 523 (2005), [astro-ph/0411247].

\bibitem{Abbott:2005kq}
LIGO Scientific, B.~Abbott {\em et~al.},
\newblock Phys. Rev. {\bf D73}, 062001 (2006), [gr-qc/0509129].

\bibitem{LIGO}
http://www.ligo.caltech.edu/.

\bibitem{Hewitson:2007zza}
LIGO Scientific, M.~Hewitson,
\newblock Class. Quant. Grav. {\bf 24}, S445 (2007).

\bibitem{GEO}
http://www.geo600.org/.

\bibitem{Acernese:2008zzf}
F.~Acernese {\em et~al.},
\newblock Class. Quant. Grav. {\bf 25}, 184001 (2008).

\bibitem{Ando:2001ej}
TAMA, M.~Ando {\em et~al.},
\newblock Phys. Rev. Lett. {\bf 86}, 3950 (2001), [astro-ph/0105473].

\bibitem{Danzmann:2003tv}
K.~Danzmann and A.~Rudiger,
\newblock Class. Quant. Grav. {\bf 20}, S1 (2003).

\bibitem{Blanchet2006}
L.~Blanchet,
\newblock Living Rev.\ Rel.\ {\bf 4} (2006),
\newblock http://www.livingreviews.org/lrr-2006-4, cited on 28 April 2011.

\bibitem{Berti:2009kk}
E.~Berti, V.~Cardoso and A.~O. Starinets,
\newblock Class. Quant. Grav. {\bf 26}, 163001 (2009), [0905.2975].

\bibitem{Arnowitt:1962hi}
R.~L. Arnowitt, S.~Deser and C.~W. Misner,
\newblock gr-qc/0405109.

\bibitem{York1979}
J.~W. {York}, Jr.,
\newblock {Kinematics and dynamics of general relativity},
\newblock in {\em Sources of Gravitational Radiation}, edited by {L.~L.~Smarr},
  pp. 83--126, 1979.

\bibitem{Pretorius:2005gq}
F.~Pretorius,
\newblock Phys. Rev. Lett. {\bf 95}, 121101 (2005), [gr-qc/0507014].

\bibitem{Baker:2005vv}
J.~G. Baker, J.~Centrella, D.-I. Choi, M.~Koppitz and J.~van Meter,
\newblock Phys. Rev. Lett. {\bf 96}, 111102 (2006), [gr-qc/0511103].

\bibitem{Campanelli:2005dd}
M.~Campanelli, C.~O. Lousto, P.~Marronetti and Y.~Zlochower,
\newblock Phys. Rev. Lett. {\bf 96}, 111101 (2006), [gr-qc/0511048].

\bibitem{Pretorius:2007nq}
F.~Pretorius,
\newblock 0710.1338.

\bibitem{Alcubierre:2008}
M.~Alcubierre,
\newblock {\em {Introduction to 3+1 numerical relativity }}{International
  series of monographs on physics} (Oxford Univ. Press, Oxford, 2008).

\bibitem{Hannam:2009rd}
M.~Hannam,
\newblock Class. Quant. Grav. {\bf 26}, 114001 (2009), [0901.2931].

\bibitem{Hinder:2010vn}
I.~Hinder,
\newblock Class. Quant. Grav. {\bf 27}, 114004 (2010), [1001.5161].

\bibitem{Centrella:2010mx}
J.~M. Centrella, J.~G. Baker, B.~J. Kelly and J.~R. van Meter,
\newblock 1010.5260.

\bibitem{Dain:2008ck}
S.~Dain, C.~O. Lousto and Y.~Zlochower,
\newblock Phys. Rev. {\bf D78}, 024039 (2008), [0803.0351].

\bibitem{Sperhake:2008ga}
U.~Sperhake, V.~Cardoso, F.~Pretorius, E.~Berti and J.~A. Gonzalez,
\newblock Phys. Rev. Lett. {\bf 101}, 161101 (2008), [0806.1738].

\bibitem{Shibata:2008rq}
M.~Shibata, H.~Okawa and T.~Yamamoto,
\newblock Phys. Rev. {\bf D78}, 101501 (2008), [0810.4735].

\bibitem{Sperhake:2009jz}
U.~Sperhake {\em et~al.},
\newblock Phys. Rev. Lett. {\bf 103}, 131102 (2009), [0907.1252].

\bibitem{Hannam:2006vv}
M.~Hannam, S.~Husa, D.~Pollney, B.~Bruegmann and N.~O'Murchadha,
\newblock Phys. Rev. Lett. {\bf 99}, 241102 (2007), [gr-qc/0606099].

\bibitem{Brown:2007tb}
J.~D. Brown,
\newblock Phys. Rev. {\bf D77}, 044018 (2008), [0705.1359].

\bibitem{Brown:2007nt}
J.~D. Brown,
\newblock Class. Quant. Grav. {\bf 25}, 205004 (2008), [0705.3845].

\bibitem{Hannam:2008sg}
M.~Hannam, S.~Husa, F.~Ohme, B.~Bruegmann and N.~O'Murchadha,
\newblock Phys. Rev. {\bf D78}, 064020 (2008), [0804.0628].

\bibitem{Brugmann:2009gc}
B.~Bruegmann,
\newblock Gen. Rel. Grav. {\bf 41}, 2131 (2009), [0904.4418].

\bibitem{Shibata:1995we}
M.~Shibata and T.~Nakamura,
\newblock Phys. Rev. {\bf D52}, 5428 (1995).

\bibitem{Baumgarte:1998te}
T.~W. Baumgarte and S.~L. Shapiro,
\newblock Phys. Rev. {\bf D59}, 024007 (1999), [gr-qc/9810065].

\bibitem{Laguna:2002zc}
P.~Laguna and D.~Shoemaker,
\newblock Class. Quant. Grav. {\bf 19}, 3679 (2002), [gr-qc/0202105].

\bibitem{Will:2005va}
C.~M. Will,
\newblock Living Rev.\ Rel.\ {\bf 9}, 3 (2006), [gr-qc/0510072],
\newblock http://www.livingreviews.org/lrr-2006-3, cited on 28 April 2011.

\bibitem{Yunes:2009hc}
N.~Yunes and F.~Pretorius,
\newblock Phys. Rev. {\bf D79}, 084043 (2009), [0902.4669].

\bibitem{Salgado:2008xh}
M.~Salgado, D.~M.-d. Rio, M.~Alcubierre and D.~Nunez,
\newblock Phys. Rev. {\bf D77}, 104010 (2008), [0801.2372].

\bibitem{Antoniadis:1990ew}
I.~Antoniadis,
\newblock Phys. Lett. {\bf B246}, 377 (1990).

\bibitem{ArkaniHamed:1998rs}
N.~Arkani-Hamed, S.~Dimopoulos and G.~R. Dvali,
\newblock Phys. Lett. {\bf B429}, 263 (1998), [hep-ph/9803315].

\bibitem{Antoniadis:1998ig}
I.~Antoniadis, N.~Arkani-Hamed, S.~Dimopoulos and G.~R. Dvali,
\newblock Phys. Lett. {\bf B436}, 257 (1998), [hep-ph/9804398].

\bibitem{Randall:1999ee}
L.~Randall and R.~Sundrum,
\newblock Phys. Rev. Lett. {\bf 83}, 3370 (1999), [hep-ph/9905221].

\bibitem{Randall:1999vf}
L.~Randall and R.~Sundrum,
\newblock Phys. Rev. Lett. {\bf 83}, 4690 (1999), [hep-th/9906064].

\bibitem{Maldacena:1997re}
J.~M. Maldacena,
\newblock Adv. Theor. Math. Phys. {\bf 2}, 231 (1998), [hep-th/9711200].

\bibitem{Gubser:1998bc}
S.~S. Gubser, I.~R. Klebanov and A.~M. Polyakov,
\newblock Phys. Lett. {\bf B428}, 105 (1998), [hep-th/9802109].

\bibitem{Witten:1998qj}
E.~Witten,
\newblock Adv. Theor. Math. Phys. {\bf 2}, 253 (1998), [hep-th/9802150].

\bibitem{Yoshino:2009xp}
H.~Yoshino and M.~Shibata,
\newblock Phys. Rev. {\bf D80}, 084025 (2009), [0907.2760].

\bibitem{Shibata:2009ad}
M.~Shibata and H.~Yoshino,
\newblock Phys. Rev. {\bf D81}, 021501 (2010), [0912.3606].

\bibitem{Shibata:2010wz}
M.~Shibata and H.~Yoshino,
\newblock Phys. Rev. {\bf D81}, 104035 (2010), [1004.4970].

\bibitem{Zilhao:2010sr}
M.~Zilhao {\em et~al.},
\newblock Phys. Rev. {\bf D81}, 084052 (2010), [1001.2302].

\bibitem{Witek:2010xi}
H.~Witek {\em et~al.},
\newblock Phys. Rev. {\bf D82}, 10 (2010), [1006.3081].

\bibitem{Sorkin:2009bc}
E.~Sorkin and M.~W. Choptuik,
\newblock Gen. Rel. Grav. {\bf 42}, 1239 (2010), [0908.2500].

\bibitem{Sorkin:2009wh}
E.~Sorkin,
\newblock Phys. Rev. {\bf D81}, 084062 (2010), [0911.2011].

\bibitem{Choptuik:2003qd}
M.~W. Choptuik {\em et~al.},
\newblock Phys. Rev. {\bf D68}, 044001 (2003), [gr-qc/0304085].

\bibitem{Lehner:2010pn}
L.~Lehner and F.~Pretorius,
\newblock Phys. Rev. Lett. {\bf 105}, 101102 (2010), [1006.5960].

\bibitem{Dennison:2010wd}
K.~A. Dennison, J.~P. Wendell, T.~W. Baumgarte and J.~Brown,
\newblock Phys.Rev. {\bf D82}, 124057 (2010), [1010.5723].

\bibitem{Witek:2010qc}
H.~Witek {\em et~al.},
\newblock Phys.Rev. {\bf D82}, 104037 (2010), [1004.4633].

\bibitem{Sperhake:2006cy}
U.~Sperhake,
\newblock Phys. Rev. {\bf D76}, 104015 (2007), [gr-qc/0606079].

\bibitem{Beyer:2004sv}
H.~R. Beyer and O.~Sarbach,
\newblock Phys. Rev. {\bf D70}, 104004 (2004), [gr-qc/0406003].

\bibitem{Alcubierre:2002kk}
M.~Alcubierre {\em et~al.},
\newblock Phys. Rev. {\bf D67}, 084023 (2003), [gr-qc/0206072].

\bibitem{Nagy:2004td}
G.~Nagy, O.~E. Ortiz and O.~A. Reula,
\newblock Phys. Rev. {\bf D70}, 044012 (2004), [gr-qc/0402123].

\bibitem{Gundlach:2006tw}
C.~Gundlach and J.~M. Martin-Garcia,
\newblock Phys. Rev. {\bf D74}, 024016 (2006), [gr-qc/0604035].

\bibitem{Garfinkle:2007yt}
D.~Garfinkle, C.~Gundlach and D.~Hilditch,
\newblock Class. Quant. Grav. {\bf 25}, 075007 (2008), [0707.0726].

\bibitem{Bernuzzi:2009ex}
S.~Bernuzzi and D.~Hilditch,
\newblock Phys. Rev. {\bf D81}, 084003 (2010), [0912.2920].

\bibitem{Calabrese:2005ft}
G.~Calabrese, I.~Hinder and S.~Husa,
\newblock J. Comput. Phys. {\bf 218}, 607 (2006), [gr-qc/0503056].

\bibitem{Chirvasa:2008xx}
M.~Chirvasa and S.~Husa,
\newblock 0812.3752.

\bibitem{Gustafsson1995}
B.~Gustafsson, H.~O. Kreiss and J.~Oliger,
\newblock {\em {Time dependent problems and difference methods}} (Wiley, 1995).

\bibitem{cactusweb}
{Cactus} {Computational} {Toolkit},
\newblock http://www.cactuscode.org/.

\bibitem{carpetweb}
Mesh refinement with {Carpet},
\newblock http://www.carpetcode.org/.

\bibitem{Schnetter:2003rb}
E.~Schnetter, S.~H. Hawley and I.~Hawke,
\newblock Class. Quant. Grav. {\bf 21}, 1465 (2004), [gr-qc/0310042].

\bibitem{Ansorg:2004ds}
M.~Ansorg, B.~Bruegmann and W.~Tichy,
\newblock Phys. Rev. {\bf D70}, 064011 (2004), [gr-qc/0404056].

\bibitem{Alcubierre:2003pc}
M.~Alcubierre {\em et~al.},
\newblock Class. Quant. Grav. {\bf 21}, 589 (2004), [gr-qc/0305023].

\bibitem{Baker:2006yw}
J.~G. Baker, J.~Centrella, D.-I. Choi, M.~Koppitz and J.~van Meter,
\newblock Phys. Rev. {\bf D73}, 104002 (2006), [gr-qc/0602026].

\bibitem{Brugmann:2008zz}
B.~Bruegmann {\em et~al.},
\newblock Phys. Rev. {\bf D77}, 024027 (2008), [gr-qc/0610128].

\bibitem{magerit2010}
http://www.cesvima.upm.es/.

\bibitem{BSC-RES2010}
http://www.bsc.es/.

\bibitem{HD_LaSh_inprep}
H.~Witek {\em et~al.},
\newblock {work in progress}.

\end{thebibliography}

\end{document}